\documentclass[11pt, a4paper]{article}
\usepackage{jheppub}
\usepackage{graphicx}
\usepackage{epsfig,amsmath,amsthm}
\usepackage{epstopdf}

\newcommand{\be}{\begin{equation}}
\newcommand{\ee}{\end{equation}}
\newcommand{\bea}{\begin{eqnarray}}
\newcommand{\eea}{\end{eqnarray}}
\def\bse{\begin{subequations}}
\def\ese{\end{subequations}}

\def\IZ{\relax\ifmmode\hbox{Z\kern-.4em Z}\else{Z\kern-.4em Z}\fi}

\newcommand{\non}{\nonumber \\}

\def\del{{\partial}}

\def\hd{{\hat d}} 
 
\def\hS{{\hat S}}

 \def\co{{\cal O}}

\def\al{\alpha} 
\def\gm{\gamma}  \def\eps{\epsilon}

\def\bi{\begin{itemize}} \def\ei{\end{itemize}}

\def\({\left(} \def\){\right)}
\def\[{\left[} \def\]{\right]}

\def\w{\omega}
\def\Om{\Omega}

\def\PhiS{A_E}			
\def\PhiV{A_{M\aleph}}		
\def\REone{R^+_1}
\def\RMone{R^-_1}
\def\d{\partial}

\def\Omd{\Om_{\hat{d}+1}}
\def\dOmd{d\Omd}
\def\Gm{\Gamma}
\def\elld{_{\ell,\hd}}
\def\Nld{N\elld}
\def\Mld{M\elld}
\def\Valph{{V\aleph}}
\def\AlOm{{\aleph \, \Om}}
\def\IG{\mathbb{G}_\al}
\def\IGn{\mathbb{G}_n}
\def\tldja{\tilde{j}_{\alpha}}
\def\tldya{\tilde{y}_{\alpha}}
\def\tldhap{\tilde{h}^+_{\alpha}}

\title{An action for reaction in general dimension}

\author{Ofek Birnholtz and Shahar Hadar\\
{\it Racah Institute of Physics, Hebrew University, Jerusalem 91904, Israel} \\
{\tt ofek.birnholtz@mail.huji.ac.il}, {\tt shaharhadar@phys.huji.ac.il}
}

\abstract{We present an effective field theory study of radiation and radiation reaction effects for scalar and electromagnetic fields in general spacetime dimensions.
Our method unifies the treatment of outgoing radiation and its reaction force within a single action principle.
Central ingredients are the field doubling method, which is the classical version of the closed time path formalism and allows a treatment of non-conservative effects within an action, action level matching of system and radiation zones, and the use of fields which are adapted to the enhanced symmetries of each zone.
New results include compact expressions for radiative multipoles, radiation, and radiation reaction effective action in any spacetime dimension.
We emphasize dimension-dependent features such as the difference between electric and magnetic multipoles in higher dimensions and the temporal non-locality nature of the effective action for odd spacetime dimensions, which is a reflection of indirect propagation ("tail effect") in the full theory.
This work generalizes the method and results developed for 4 dimensions in [arXiv:1305.6930], and prepares the way for a treatment of the gravitational case.
}

\begin{document}
\maketitle

\section{Introduction}
 \label{section:Intro}
While extensive work has been done on the post-Newtonian (PN) 2-body problem in General Relativity (GR) in four spacetime dimensions (see section (1) of paper I \cite{BirnholtzHadarKol2013a} for a comprehensive review of existing literature), much less is known in general spacetime dimensions ($d$).
In this paper we study PN radiative effects, namely radiation and radiation reaction (RR), for (massless) scalar and electromagnetic (EM) theories in general $d$.
We work within an effective field theory (EFT) approach, which was introduced in \cite{GoldbergerRothstein1} to the PN limit of GR and applied to the study of dissipative effects in 4-dimensional GR also in \cite{GoldbergerRoss, GalleyLeibovich, GalleyNonConservative, GalleyTiglio, GoldbergerRothstein2, FoffaSturani4PNa}.
Our study of non-conservative effects in scalar and EM theories gives new results and insights into the problem already in these physically important cases, and lays foundations for a comprehensive treatment of these effects in higher $d$ GR (see \cite{Cardoso:2008gn} for the first treatment of gravitational radiation in higher $d$ within the EFT approach), where the main additional complication is the theory's nonlinearity.

The well known 4$d$ Abraham-Lorentz-Dirac (ALD) formula \cite{Abraham, Dirac, JacksonALD, Poisson:1999tv} gives the EM self-force on a point charge in flat spacetime in a covariant manner. There has been an extensive effort to generalize ALD to any $d$ undertaken by Kosyakov, Gal'tsov, Kazinski and others \cite{Kosyakov:1999np, Galtsov:2001iv, Kazinski:2002mp, Galtsov:2004qz, Kazinski:2004qq, Galakhov:2007my, Kosyakov:2007qc, Kosyakov:2008wa, Shuryak:2011tt}.
Joint treatment of self-force under scalar, vector and tensor fields has been given by \cite{Mironov:2007nk,Mironov:2007mv,Galtsov:2007zz, Galtsov:2010cz}.
As is well known, in even $d$ waves exhibit \emph{direct} propagation, that is the Green's function is supported only on the lightcone, while in non-even $d$ waves propagate also \emph{indirectly} - exhibiting the so-called tail effect.
These features appear also in the above treatments of RR, but to our knowledge a closed-form formula for generalized ALD is available currently only for EM in even $d$ \cite{Kazinski:2002mp}. The very regularizability of RR in high $d$ is sometimes put to question in the above treatments.

\begin{figure}[t!]
\centering \noindent
\includegraphics[width=15cm]{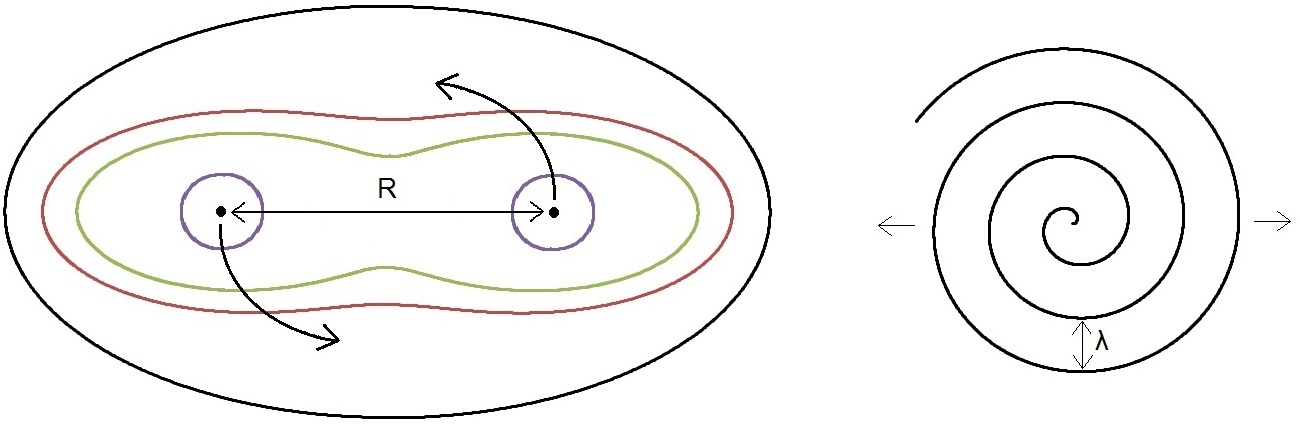}
\caption[]{Schematic sketch of the two relevant zones.
On the left side is the system zone, with a typical stationary-like field configuration.
On the right side is the radiation zone with its typical out-spiraling waves.}
 \label{fig:zones}
\end{figure}

Our method divides the problem into 2 zones, the system and radiation zones (fig.\ref{fig:zones}).
Each zone has an enhanced symmetry, namely a symmetry that is not a symmetry of the full problem. Different symmetries emerge when zooming out to the radiation zone or in to the system zone.
The system zone is approximately stationary (time independent) since by assumption (of the PN approximation) all velocities are non-relativistic.
The radiation zone is approximately spherically symmetric since the system shrinks to a point at the origin and hence rotations leave it invariant.
Identifying symmetries is central to making fitting choices for the formulation of a perturbation theory in each zone, including the choice of how to divide the action into a dominant part and a perturbation, the choice of field variables and - when relevant - the choice of gauge.
Hence we insist on using spherical field variables (in fact, symmetric trace-free tensors), following paper I and \cite{AsninKol}, and unlike the more common plane-wave and wave-vectors approach used in many EFT works \cite{GoldbergerRothstein2,GoldbergerRoss, GalleyTiglio,GalleyLeibovich, FoffaSturani4PNa}.

As is well known, Hamilton's traditional action formalism is not compatible with dissipative effects.
In order to account for non-conservative effects within the action we use the method of field doubling (see \cite{GalleyNonConservative} for a crisp general formulation) which is the classical version of the closed time path (CTP) or in-in formalism \cite{CTP} introduced in the quantum field theory context in the 1960's.
In this method each degree of freedom in the system is \emph{doubled} and a generalized action principle for the whole system (including doubled fields) is constructed from the original action.
Only after deriving equations of motion (EOM) from this generalized action, one enforces the so-called physical condition, identifying each field with its doubled counterpart.
The resulting forces in the EOM can be dissipative.

In order to formulate the whole problem within a single action principle, we introduce new matching fields which couple to system \& radiation zone fields at an appropriate boundary.
We call these fields ``2-way multipoles".
We physically interpret them as the multipole moments of the system (living at the origin of the radiation zone and at infinity of the system zone), but they are treated as any other field in the theory.
Thereby, matching is lifted to the level of the action.

The {\bf \emph{main goal}} of this paper is to compute, in general $d$, the radiative multipoles, outgoing radiation, RR effective action, dissipated energy and RR force in the cases of scalar and electromagnetic fields, thereby generalizing the method devised in paper I to general $d$.
We comment that this paper is not intended to be self contained: the main ideas and ingredients of our formulation are thoroughly elaborated on in paper I, and here we generalize and apply them to obtain new results in higher dimensions.
The ability to generalize the machinery set up in paper I in order to obtain new results demonstrates the method's efficiency.
Moreover, this generalization is natural since a central step in the method is a reduction to one dimension (corresponding to the radial coordinate) - and from there on the treatment is very similar to the one done in 4$d$.

This paper is organized as follows:
In section \ref{section:Scalar SF} we treat the scalar field case, which already contains most of our formulation's main ingredients.
In particular this section contains a thorough discussion of the tail effect in non-even spacetime dimensions.
In section \ref{section:Electromagnetism} we treat the EM case, where one must account for the additional polarizations available for general dimensional vector fields.
In both sections we also show explicitly successful comparisons to known results in 4$d$ and 6$d$.
Sections \ref{section:Scalar SF}, \ref{section:Electromagnetism} contain detailed derivations of our results, intended to maximally clarify our computations for interested readers.
In section \ref{section:Summary of results} we briefly summarize our main results and definitions.
In section \ref{section:Discussion} we discuss these results and elaborate on future directions.

{\bf Conventions and nomenclature}: We use the mostly plus signature for the flat d-dimensional spacetime metric $\eta_{\mu\nu}$, as well as $c=1$.
We denote $D:=d-1$, $\hat{d}:= d-3$, and $\Omd$ is the volume of a unit $\hat{d}+1$ dimensional sphere.
Lower case Greek letters denote $\{0,1,..,D\}$ spacetime indices, lower case Latin letters denote $\{1..D\}$ spatial indices, upper case Greek letters denote $\{1..(\hd+1)\}$ indices on the sphere, upper Latin letters are spatial multi-indices, and Hebrew letters ($\aleph$) enumerate different vectorial harmonics on the sphere.\\

{\bf Field doubling conventions}: We work in the Keldysh (\cite{CTP}) representation of the field doubling formalism, where for every field $\phi$ (and source function) in the original action, we introduce a counterpart $\hat{\phi}$, interpreted as the difference between the doubled degrees of freedom, and with it a new EOM:
\bea
\frac{\delta \, \hat{S}}{\delta \, \hat\phi} \, = \, 0 \, \, .
\label{ctp intro}
\eea
$\hat{S}$ is always linear in the hatted fields, so enforcing the so-called physical condition $\hat\phi \, = \, 0$ (after the derivation of EOM) is trivial.

We write the radiation reaction (RR) effective action $\hat{S}$ in terms of the system's multipoles and their doubled counterparts, which are defined as
\bea
\hat{Q} =  \frac{\delta Q}{\delta \rho} \hat{\rho} \, \, .
\label{sum: doubled multipoles}
\eea
In particular, for a source composed of point particles the doubled multipoles are given by
\bea
\hat{Q} = \sum_A \frac{\delta Q}{\delta x_A} \hat{x}_A \, \, ,
\label{sum: doubled multipoles pp}
\eea
and the EOM for the $A^{th}$ point particle is $\frac{\delta \hat{S}}{\delta \hat{x}_A} \, = \, 0 \, \, $
 For a detailed survey of the doubling formalism used in this paper, see section (2.4) of paper I.

\section{Scalar case}
 \label{section:Scalar SF}
\subsection{Spherical waves and double-field action}
 \label{spherical waves and double-field action}
We start with a massless scalar field coupled to some charge distribution in arbitrary spacetime dimension $d$.
We take the action to be
\bea
S_\Phi= +\frac{1}{2 \Omd\, G} \int\! (\d_\mu \Phi)^2 r^{\hat{d}+1} dr \dOmd dt  - \int\!\! \rho\Phi r^{\hat{d}+1} dr \dOmd dt\, , \,\,\,\,\,\,\,\,
\label{Scalar Action}
\eea
which leads to the usual field equation
\bea
\Box \phi = -\Omd G \rho \, .
\label{Scalar Field Equation}
\eea
{\bf Spherical waves: conventions}.
We shall work in the frequency domain and use a basis of spherical decomposition to multipoles.
We note that any symmetric trace free tensor $\phi_{L_\ell}$ ($\ell$ being given), can be represented equivalently using the standard (scalar) spherical harmonic representation $\phi_{\ell m}$
(who form a basis of dimension $D_\ell(\hd+1,0)$\footnote{$D_\ell(n,s)$ is the number of independent spherical harmonics of degree $\ell$ and spin $s$ on the $n$-sphere. We use $D_\ell(\hd+1,0)=\frac{(2\ell+\hd)(\ell+\hd-1)!}{\ell!\hd!}$ \cite{RubinOrdonez,Higuchi:1986wu}.}) or using functions on the unit sphere $\phi_\ell(\Omd)$, with the different forms related by
\bea 	
\phi_\ell(\Omd) = \phi_{L_\ell}\, \frac{x^{L_\ell}}{r^\ell} = \sum_{m} \phi_{\ell m}\, Y_{\ell m}(\Omd).
\label{multipole basis changes}
\eea
We found it convenient to use the $\phi_L$ decomposition, which is similar to the Maxwell cartesian spherical multipoles \cite{Maxwell,FryeEfthimiou,Kalf,Applequist}.
We thus decompose the field and the sources as
\bea
\label{decomposition of scalar field}
\Phi(\vec{r},t)&=&\int\frac{d \w}{2\pi}\sum_L e^{-i \w t} \Phi_{L \w}(r) x^L \, , \\
\rho({\vec{r}},t)&=&\int\frac{d \w}{2\pi}\sum_L e^{-i \w t} \rho_{L \w}(r) x^L \, ,
\label{decomposition of scalar field and source}
\eea
where $L=(k_1k_2 \cdots k_\ell)$ is a multi-index and $x^L$ is the corresponding symmetric-trace-free (STF) multipole
\bea
x^L=(x^{k_1}x^{k_2} \cdots x^{k_\ell})^{STF} \equiv r^\ell n^L.
\label{nL STF}
\eea
With $g^{\Om \Om'}$ the metric on the $\hat{d}+1$ dimensional unit sphere, the $x^L$ are eigenfunctions of the Laplacian operator on that sphere,
\bea
\Delta_{\Omd} x^L=-c_s x^L = -\ell(\ell+\hat{d}) x^L.
\label{xL eigen-value}
\eea
For any dimension d, an orthogonal basis can be constructed from the multipoles, satisfying
\bea
\int x_{L_\ell}(r,\Omd) x^{L'_{\ell '}}(r,\Omd) \dOmd
	&=&\Nld r^{2\ell} \Omd \delta_{\ell \ell'} \delta_{L_\ell L'_{\ell'}} ,
\label{scalar spherical harmonics normalization} \\
\int g^{\Om \Om'} \partial_{\Om} x_{L_\ell}(r,\Omd) \partial_{\Om'} x^{L'_{\ell '}}(r,\Omd) \dOmd
	&=&c_s \cdot \Nld r^{2\ell} \Omd \delta_{\ell \ell'} \delta_{L_\ell L'_{\ell'}} ,
\label{scalar spherical harmonics normalization 2}	\\
\Nld=\frac{\Gm(1+\hd/2)} {2^\ell\,\Gm(\ell+1+\hd/2)}=\frac{\hd!!}{(2\ell+\hd)!!}&.
\label{Nld}
\eea
We use the summation conventions and definitions following \cite{BirnholtzHadarKol2013a} (see Appendix \ref{app:Multi-index summation convention}) and note that $d=4$ implies $\hd=1$, $N_{\ell,1}=(2\ell+1)!!^{-1}$.
We also use the inverse transformation
\bea
\rho_{L\w}(r)  &=&
\!\int\!\! \rho_\w(\vec{r}) x_L \frac{\dOmd}{\Nld\Omd r^{2\ell}}=
\int\!\!\!\!\int\!\!{dt}e^{i \w t} \rho(\vec{r},t) x_L \frac{\dOmd}{\Nld\Omd r^{2\ell}} \, .
\label{scalar inverse source}
\eea
{\bf Spherical waves: dynamics}.
In the new notation, using $\Phi_{L -\w}=\Phi^*_{L \w}$, the action (\ref{Scalar Action}) becomes
\bea
S_\Phi&=&\frac{1}{2}\!\int\!\!\frac{d \w}{2\pi}\!\sum_L \!\!\int\!\!{dr}\!
\[ \frac{r^{2\ell+\hd+1} \Nld }{\, G}
\Phi^*_{L \w} \( \w^2+\d_r^2+\frac{2\ell+\hd+1}{r}\d_r \) \Phi_{L \w}
-\( \rho^\Phi_{L \w}\Phi^*_{L \w}+c.c.\) \]  \!, \non
\label{scalar action spherical}
\eea
with the source term defined as
\bea
\rho_{L \w}^\Phi (r)= \Nld \, \Omd r^{2\ell+\hd+1}\rho_{L \w}(r)=r^{\hd+1} \int{\dOmd}\rho_\w(\vec{r}) x_L.
\label{scalar action source}
\eea
From (\ref{scalar action spherical}) we derive the EOM
\bea
0=\frac{\delta S}{\delta \Phi_{L\w}^*}=\frac{\Nld \, r^{2\ell+\hd+1} }{G}  \( \w^2+\d_r^2+\frac{2\ell+\hd+1}{r}\d_r  \)  \Phi_{L \w}-\rho^\Phi_{L \w} \, \, .
\label{EOM Phi}
\eea
Defining $x:= \w r$ the homogenous part of this equation is
\bea
[ \del_x^2 + \frac{2\ell+\hd+1}{x}\del_x + 1]\, \psi\elld=0,
\label{Modified Bessel equation}
\eea
and its solutions $\psi\elld=\psi_\alpha=\tilde{j}_\alpha,\,\tilde{y}_\alpha,\,\tilde{h}^\pm_\alpha$ (with $\alpha=\ell+\hd/2$) are Bessel functions up to normalization (see Appendix \ref{app:Normalizations of Bessel functions}).
Thus \emph{the propagator for spherical waves is}
\bea
\parbox{20mm}{\includegraphics[scale=0.5]{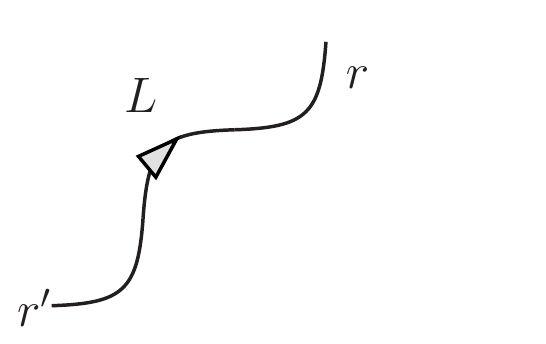}}	
 \equiv G^\Phi_{ret}(r',r)
= -i \w^{2\ell+\hd} \Mld \, G \, \tilde{j}_{\alpha}(\w r_1)\, \tilde{h}^\pm_{\alpha}(\w r_2) \, \, ;
\label{Phi propagator scalar} \\
r_1:=\min\{r',r\}, ~ r_2:=\max\{r',r\} \, , \nonumber
\eea
where
\bea
\Mld = \frac{\pi}{2^{2\alpha+1}\Nld \, \Gm^2(\alpha+1)}=
\frac{\pi}{2^{\ell+1+\hd}\Gm(1+\hd/2) \, \Gm(\ell+1+\hd/2)},
\label{Phi propagator scalar normalization}
\eea
which we notice is equal to $ \[ \hd!!(2\ell+\hd)!! \] ^{-1}$ for odd $\hd$ and to $\frac{\pi}{2} \[ \hd!!(2\ell+\hd)!! \] ^{-1}$ for even $\hd$.
We turn to derive the source terms (vertices) in the radiation zone. This is done through matching with the system zone according to the diagrammatic definition
\bea
\includegraphics[scale=0.5]{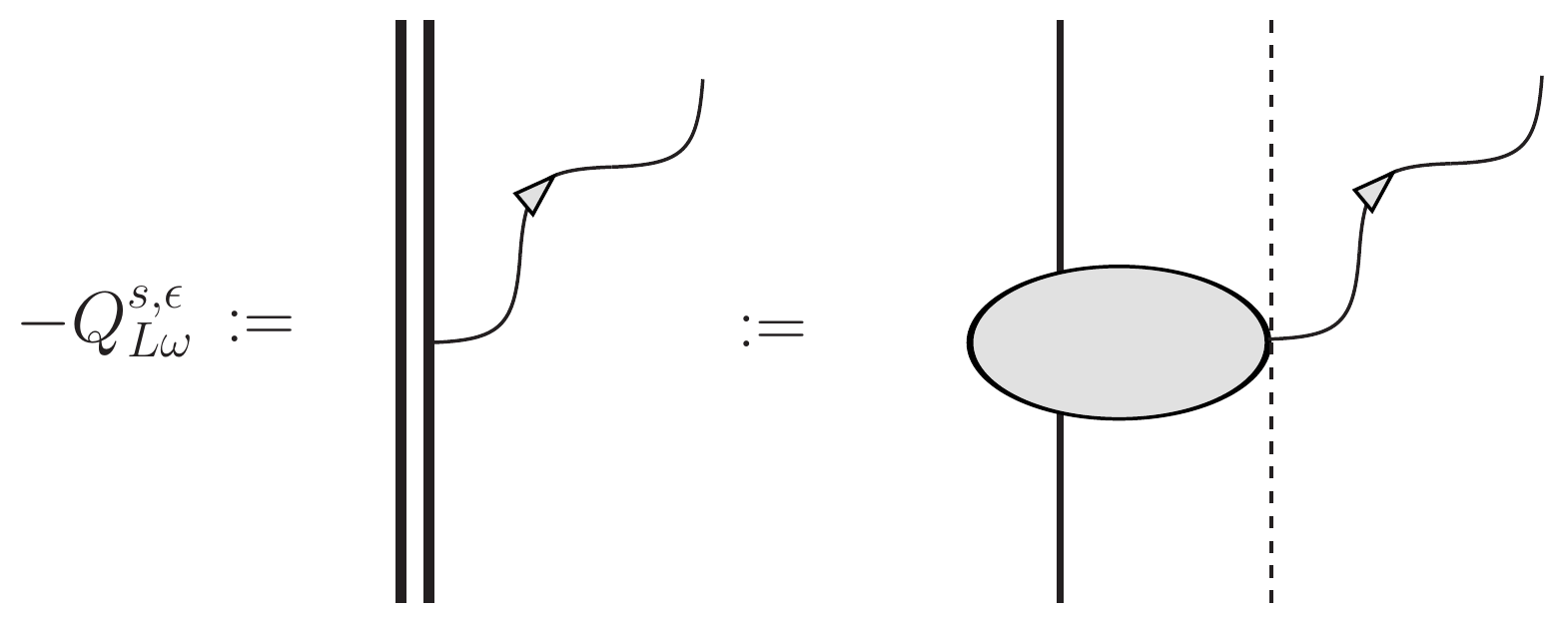} \, \, \, \, .
\label{vertex_definition}
\eea
From the radiation zone point of view the sources $Q_{L \w}$ as located at the origin or $r=0$.
Hence the radiation zone field can be written as
\bea
\Phi_{L \w}^{EFT}(r) =
\parbox{20mm} {\includegraphics[scale=0.5]{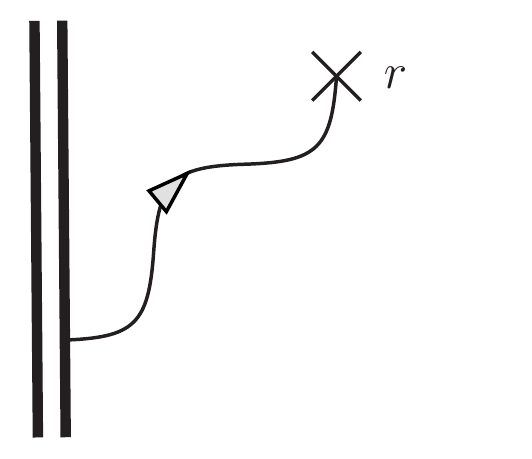}}
= - Q_{L \w}  \( -i G \w^{2\ell+\hd} \Mld  \)  \tilde{h}^+_{\alpha}(\w r) \, \, .
\label{scalar wavefunction at radiation zone1}
\eea
In the full theory (or equivalently in the system zone), on the other hand, we can also use spherical waves to obtain the field outside the source as
\bea
\Phi_{L \w}(r) &=& -\int dr' \rho^{\Phi}_{L'\w}(r') G^\Phi_{ret}(r',r)
	= -\,  \[  \int dr' \tilde{j}_{\alpha}(\w r') \rho_{L \w}^\Phi (r')   \]    \( -i G \w^{2\ell+\hd} \Mld  \)  \tilde{h}^+_{\alpha}(\w r) \nonumber \\
 &=& -\,  \[  \int d^D x' \tilde{j}_{\alpha}(\w r') \rho_\w(\vec{r} \, ') x'_L   \]    \( -i G \w^{2\ell+\hd} \Mld  \)  \tilde{h}^+_{\alpha}(\w r) \, \, .
\label{scalar wavefunction at radiation zone2}
\eea
By comparing the above expressions for the field (\ref{scalar wavefunction at radiation zone1},\ref{scalar wavefunction at radiation zone2}) and using (\ref{scalar inverse source}) to return to the time domain we find that \emph{the radiation source multipoles are}
\be
Q_L = \int d^D x\, \tilde{j}_\alpha(ir\d_t)\, x^{STF}_L\, \rho(\vec{r},t) ~ .
\label{I Phi scalar multipoles}
\ee
We note that from the series expansion of $\tilde{j}_\alpha$ (\ref{Bessel J series2}) it can be seen that $Q^L$ includes only even powers of $i\d_t$, for every $\ell,\hd$, and is thus well-defined and real.
A simple test of substituting $d=4$ shows these multipoles coincide with the multipoles given by Ross \cite{RossMultipoles} and by paper I.
We can think of this process as a ``zoom out balayage'' (French for sweeping/scanning - see paper I) of the original charge distribution $\rho({\vec{r}})$ into $Q_L$ carried out through propagation with $\tilde{j}_{\alpha}(\w r)$.
A useful representation of this result is
\bea
Q_L = \int d^D x \, x^{STF}_L \int_{-1}^{1}dz \delta_{\elld}(z) \rho(\vec{r},u+z r) \, ~ ,
\label{I Phi scalar multipoles using delta ell}
\eea
where (inspired by \cite{Blanchet:1989ki,Damour:1990gj}) we have implicitly defined the $\hd$-dimensional generating time-weighted function
\bea
\int_{-1}^{1}dz \delta_{\elld}(z) f(\vec{r},u+z r)&=&\sum_{p=0}^{+\infty}\frac{(2\ell+\hd)!!}{(2p)!!(2\ell+2p+\hd)!!}  \(  r\d_u  \) ^{2p} f(\vec{r},u)~,
\label{generating funtion delta ell}
\eea
remarking that for odd $\hd$ (even space-time dimension),
\bea
\delta_{\elld}(z)&=&\frac{(2\ell+\hd)!!}{2 (2\ell+\hd-1)!!}(1-z^2)^{\ell+(\hd-1)/2}~.
\label{generating funtion delta ell even}
\eea
Altogether \emph{the Feynman rules in the radiation zone} for the propagator and vertices are
\begin{align}
\label{Feynman Rule Vertex}
\parbox{20mm} {\includegraphics[scale=0.3]{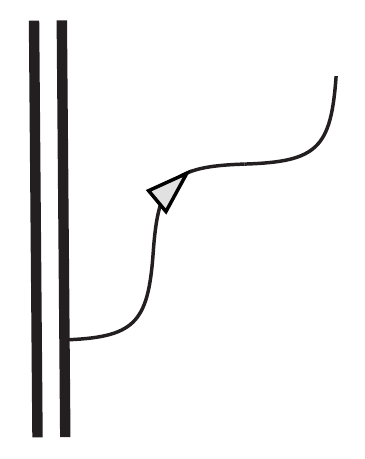}}
=-Q^{s,\epsilon}_{L \w}\,\,\,\,\,\,\,\,\,\, ,\,\,\,\,\,
\parbox{20mm} {\includegraphics[scale=0.3]{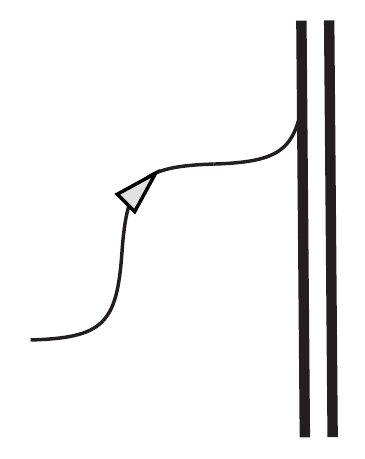}}
=-\hat{Q}^{s,\epsilon *}_{L \w}\,\,\,\, ,
\end{align}
\begin{align}
\label{Feynman Rule Propagator}
\parbox{20mm} {\includegraphics[scale=0.5]{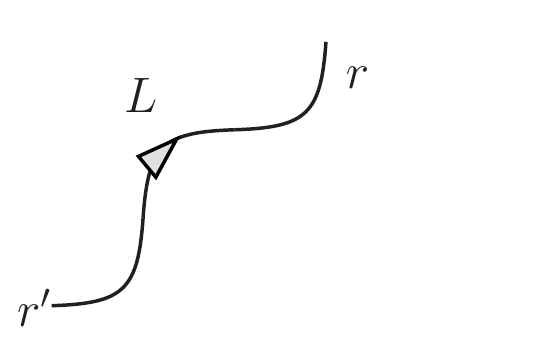}}
=G^{s,\epsilon}_{ret}(r',r)=
-i \w^{2\ell+\hd} \Mld \, R_s^\epsilon \,G \, \tilde{j}_{\alpha}(\w r_1)\, \tilde{h}^+_{\alpha}(\w r_2)
\,\,\,\,\,\,\,\,\,\,\,r_1  \leq r_2	 \, \, ,
\end{align}
where for future use we allowed for a possible polarization label $\eps$ and a rational $\ell,\hd$-dependent factor $R_s^\epsilon$ which is absent in the scalar case, namely $R|_{s=0}=1$.

\subsection{Outgoing radiation and the RR effective action}
\label{Outgoing radiation and the RR effective action}
We can now use these Feynman rules to compute central quantities.\\
{\bf Outgoing radiation} can now be found diagrammatically as
\begin{align}
\label{Radiation Phi using feynman}
\Phi_{L\w}(r)=&
\parbox{20mm} {\includegraphics[scale=0.5]{DiagRadiationScalar.pdf}}
= 	-Q_{L'\w}G^\Phi_{ret}(0,r)
=\sqrt{\frac{\pi}{2^{\hd+1}}} \frac{G}{\Gm(1+\hd/2)} (-i\w)^{\ell+\frac{\hd-1}{2}} \frac{Q_{L\w}}{r^{\ell}} \frac{e^{i\w r}}{r^{\frac{\hd+1}{2}}},~~~~
\end{align}
where we used the asymptotic forms of $\tilde{h}_\alpha(x)$ (\ref{Bessel H asymptotic2}) and $\tilde{j}_\alpha(\w r')|_{r'=0}=1$ for the source at $r'=0$, and (\ref{Phi propagator scalar normalization}).
In odd $\hd$ we can use (\ref{decomposition of scalar field and source}) to find that (as $r \to \infty$) \emph{the outgoing radiation is}
\bea
\Phi(\vec{r},t) \sim \frac{G}{\hd!!} r^{-\frac{\hd+1}{2}}
\sum_L n^L \d_t^{\ell+\frac{\hd-1}{2}} Q_L(t-r) ~~  \, . ~
\label{radiation Phi}
\eea
For $d=4$ this coincides with paper I (eq.(3.20)) and with \cite{RossMultipoles} (eq. (21); note our normalizations differ by $4\pi$).
While (\ref{radiation Phi}) is valid and local for all odd $\hd$, for non-odd $\hd$ (non-even spacetime dimension), the corresponding expression in the time-domain would include a non-integer number of time derivatives, implying non-locality (see more below).
(\ref{Radiation Phi using feynman}) is, of course, valid in every dimension.\\
{\bf Dissipated power}.
The power carried away by the radiation field (\ref{Radiation Phi using feynman}) is
\bea
\dot{E} &=& \frac{1}{G}\int \dot\Phi^2 r^{\hd+1}\dOmd
=\frac{1}{G} \int\!\! \frac{d\w}{2\pi} \sum_{L,L'} \int r^{\hd+1}\dOmd \w^2 \Phi_{L\w}^* x^L x_{L'}\Phi^{L' \w} \nonumber\\
&=&\sum_L \int\!\! \frac{d\w}{2\pi} \frac{\pi G \Nld }{2^{\hd+1} \Gm^2(1+\hd/2)} \w^{2\ell+\hd+1} Q^{L\w} Q^*_{L\w}		 \nonumber\\
&=&\sum_L \int\!\! \frac{d\w}{2\pi} G\w^{2\ell+\hd+1} |Q_{L\w}|^2
\cdot
\left\{
\begin{array}{ll}
	\frac{\pi}{2} \[ \hd!! (2\ell+\hd)!! \] ^{-1} ~ & d~ odd,\\
	 \[ \hd!! (2\ell+\hd)!! \] ^{-1} ~ & d ~ even,\\
	\pi \[ 2^{\ell+\hd+1} \Gm(\ell\!+\!1\!+\!\frac{\hd}{2}) \,\Gm(1\!+\!\frac{\hd}{2}) \] ^{-1} ~ & d ~ non-integer,\\
\end{array}
\right.
\eea
where we have used (\ref{decomposition of scalar field},\ref{scalar spherical harmonics normalization},\ref{Nld}).
We note in particular that in even dimension $d$ we can find the symmetric time-domain form,
\bea
\dot{E}=\sum_{L} \frac{G}{\hd!!(2\ell+\hd)!!} \left<(\d_t^{\ell+\frac{\hd+1}{2}}Q^L )^2\right>
= \sum_{L} \frac{G}{\ell! \hd!! (2\ell+\hd)!!} \left<(\d_t^{\ell+\frac{\hd+1}{2}}Q_{k_1 k_2 \cdots k_\ell}^{STF})^2\right>\,.~~~~
\label{radiated energy scalar multipoles}
\eea
We remark that in 4d this matches (3.22) of \cite{BirnholtzHadarKol2013a} and (15) of \cite{RossMultipoles} (note normalization and summation conventions).\\
{\bf Radiation reaction effective action} encodes the RR force and is given formally by \begin{align}
\label{S Phi using feynman}
\hS_\Phi=
\parbox{20mm} {\includegraphics[scale=0.5]{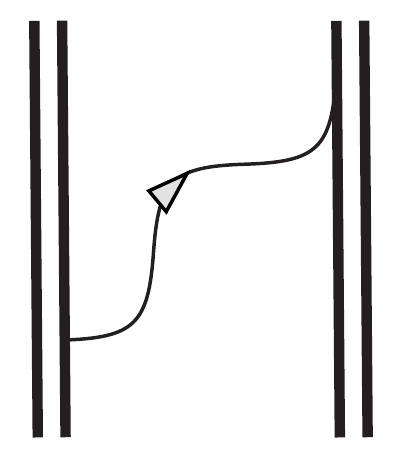}}
=& \frac{1}{2}\int\!\frac{d \w}{2\pi}\sum_{L,L'}
	 \( -Q_{L\w} \)
	G^\Phi_{ret}(0,0)
	 \( -\hat{Q}^*_{L'\w} \) +c.c.
	\nonumber\\
=& \frac{G}{2}\!\!\int\!\!\frac{d \w}{2\pi}
	\sum_{L} -i \w^{2\ell+\hd} \Mld \,
	\cdot\IG\cdot
	Q^{L\w} \hat{Q}_{L\w}^*+c.c~~,
\end{align}
\bea
\IG = \left.  \tilde{j}_{\alpha}(\w r) \right|_{r \to 0}	\cdot
		\left.  \[ \tilde{j}_{\alpha}(\w r') + i \tilde{y}_{\alpha}(\w r') \]  \right|_{r' \to 0}	~~,
\label{G definition}
\eea
where now both vertices are taken at $r=r'=0$.
For non-even $\hd$ (odd or non-integer), we use $\tilde{j}_\alpha(0)=1$, and (\ref{Bessel J series2}, \ref{Bessel Y series2}) for $\tilde{j}_\alpha,\tilde{y}_\alpha$, to find
\bea
\IG&=& \tilde{j}_{\alpha}(0) 	\cdot \tilde{j}_{\alpha}(0)
	+ \left. \left\{ \frac{i\Gm(\alpha+1)2^\alpha
			 \tilde{j}_{\alpha}(x)
			\[ J_{\alpha}(x) cos(\alpha\pi) - J_{-\alpha}(x)\]
		}{sin(\alpha\pi) x^\alpha}
	\right\} \right|_{x \to 0}	\nonumber\\
&=& \! 1 \!+\!
	\frac{i\Gm(\alpha+1)2^\alpha}{sin(\alpha\pi)}
\!\! \left.	\left\{
	\!\sum_{p,q=0}^\infty\!\! \frac{(-)^{p+q} \Gm(\alpha+1) x^{2p+2q} }{ 2^{p+q} (2p)!! (2q)!! \Gm(q\!+\!\alpha\!+\!1)}
	\! \[
		\frac{cos(\alpha\pi)}{2^\alpha \Gm(p\!+\!\alpha\!+\!1)}
		\!-\! \frac{2^\alpha x^{-2\alpha}}{\Gm(p\!-\!\alpha\!+\!1)}
	 \] \!\!
	\right\}\!\!
\right|_{x \to 0}	\!\!\!\!\!\!\!\!\!\!\!.~~~~~~~~
\label{G expression non-odd d}
\eea
The contribution from the $\tilde{j}^2_\alpha(0)=1$ term is simple and independent of dimension.
The $\tilde{j}_\alpha \tilde{y}_\alpha$ contributes both a Taylor series $x^{2p+2q}$ and a Laurent series $x^{2p+2q-2\alpha}$.
We note that for every $\hd$, the number of terms divergent as $x\to0$ is finite ($\ell+\lfloor \frac{\hd+1}{2} \rfloor$), and these amount to renormalizations of the different multipole moments.
The Taylor series consists of an infinite number of zeroes ($p+q>0$), and a single ($p=q=0$) imaginary numerical correction, $i cot(\alpha\pi)$.
Therefore for non-even $\hd$ we are left with
\bea
\IG = 1+i\cot(\pi\alpha) = i \sin^{-1}(\pi\alpha) [\cos(\pi\alpha) -i \sin(\pi\alpha) ] = [i^{2\ell+\hd-1} \sin(\pi\alpha) ]^{-1}.
\eea
Substituting in (\ref{S Phi using feynman}), we find
\bea
\hS_\Phi= \frac{G}{2}\!\!\int\!\!\frac{d \w}{2\pi}
	\sum_{L} \frac{ (-i\w)^{2\ell+\hd}}
			{ \sin(\ell\pi+\frac{\hd\pi}{2}) }
	 \Mld \,
	Q^{L\w} \hat{Q}_{L\w}^*+c.c~~,
\label{S Phi scalar multipoles general dimension}
\eea
in all non-even $\hd$.
In odd integer $\hd$, using (\ref{scalar inverse source},\ref{Phi propagator scalar normalization}) we return to the time-domain, finding \emph{the radiation reaction effective action} to be
\bea
\hS_\Phi=
G\int\!\!{dt}\sum_{L} \frac{(-)^{\ell+\frac{\hd+1}{2}}}{\hd!!(2\ell+\hd)!!}\hat{Q}^L \d_t^{2\ell+\hd}Q_L~ ,
\label{S Phi scalar multipoles}
\eea
where $Q_L$ was given by (\ref{I Phi scalar multipoles}) and
\bea
\hat{Q}^L&=&\frac{\delta Q^L}{\delta \rho}\hat{\rho}=\frac{\delta Q^L}{\delta x^{i}}\hat{x}^{i}
=\frac{\d Q^L}{\d x^{i}}\hat{x}^{i} + \frac{\d Q^L}{\d v^{i}}\hat{v}^{i} + \frac{\d Q^L}{\d a^{i}}\hat{a}^{i} + \cdots
 \, \, .
\label{I Phi scalar multipoles hat}
\eea
We remark especially that this action, and as we shall see the dissipated energy and the self force, are given by regular, local and real expressions for all odd $\hd$.
The computation itself reduces to mere multiplication: vertex -- propagator -- vertex.
The expression (\ref{S Phi scalar multipoles}) contains an odd number of time derivatives, matching our expectation of a time-asymmetric dissipative term.\\
In fractional dimensions, transforming (\ref{S Phi scalar multipoles general dimension}) back to the time domain introduces the Fourier transform of a fractional derivative, yielding
\bea
\hS_\Phi &=& \frac{G}{2} \sum_{L}
	 \frac{ \Mld } { \sin(\ell\pi+\frac{\hd\pi}{2}) }
	\int\!\!\frac{d \w}{2\pi} (-i\w)^{2\ell+\hd}
	\int\!\!dt\, e^{-i\w t} \hat{Q}^{L}(t) \int\!\!dt'\, e^{i\w t'} Q_{L}(t')+c.c \nonumber\\
&=& \frac{G}{2} \sum_{L}
	 \frac{ \Mld } { \sin(\ell\pi+\frac{\hd\pi}{2}) }
	\int\!\!dt \hat{Q}^{L}(t) \!\int\!\!dt' Q_{L}(t')
	\, \d_t^n \!\!
	\int\!\!\frac{d \w}{2\pi} (-i\w)^{-b}
	e^{i\w (t'-t)} +c.c\nonumber\\
&=& G\sum_{L}
	 \frac{ \Mld } {\Gm(b) \sin(\ell\pi+\frac{\hd\pi}{2}) }
	\int\!\!dt \hat{Q}^{L}(t) \!\int\!\!dt' Q_{L}(t')
	\, \d_t^n
	\frac{\Theta(t-t')} {(t-t')^{1-b} } \nonumber\\
&=& \frac{G}{\Gm(b)}\!\sum_{L}\!\!
	\frac{\pi}{2^{\ell+1+\hd}\Gm(1\!\!+\!\!\frac{\hd}{2}) \, \Gm(\ell\!\!+\!\!1\!\!+\!\!\frac{\hd}{2}) \sin(\ell\pi\!\!+\!\!\frac{\hd\pi}{2})} \!
	\int_{-\infty}^{\infty}\!\!\!\!\!\!\!\! dt \hat{Q}^{L}(t) \!\!\int_{-\infty}^t \!\!\!\!\!\!\!\! dt' Q_{L}(t')
	\, \d_t^n (t-t')^{b-1}~,~~~~~~~~
\eea
where $n=2\ell+\lceil\hd\rceil$, $b=\lceil\hd\rceil-\hd$.
Thus in fractional dimensions, the self-force has a non-local ("tail") contribution, originating from transforming $(-i\w)^{-b}$.
We note that this tail integral converges at $t'\!\!\to\!\! t$, and under reasonable assumptions regarding $Q^L(t)$ at very early times converges at $t'\!\!\to\!\!-\infty$ as well.
As shown earlier, in even integer dimensions $b=0$ and the transformation produces a delta function in time, indicating locality.
However, the same form does not apply at odd integer $d$, because the Bessel functions and $\IG$ assume a different form.
Using a limiting process over $d'\!=\!d\!+\!\eps$ for $\eps\!\!\to\!\!0^+$, corresponding to $n\!=\!2\ell\!+\!\hd\!+\!1$, $b\!=\!1\!-\!\eps$, we expect a $ln$ term to appear as $b\!\to\!1^-$; we treat it separately.

\subsection{Odd spacetime dimensions}
\label{Odd spacetime dimensions}
\begin{figure}[t!]
\centering \noindent
\includegraphics[scale=0.3]{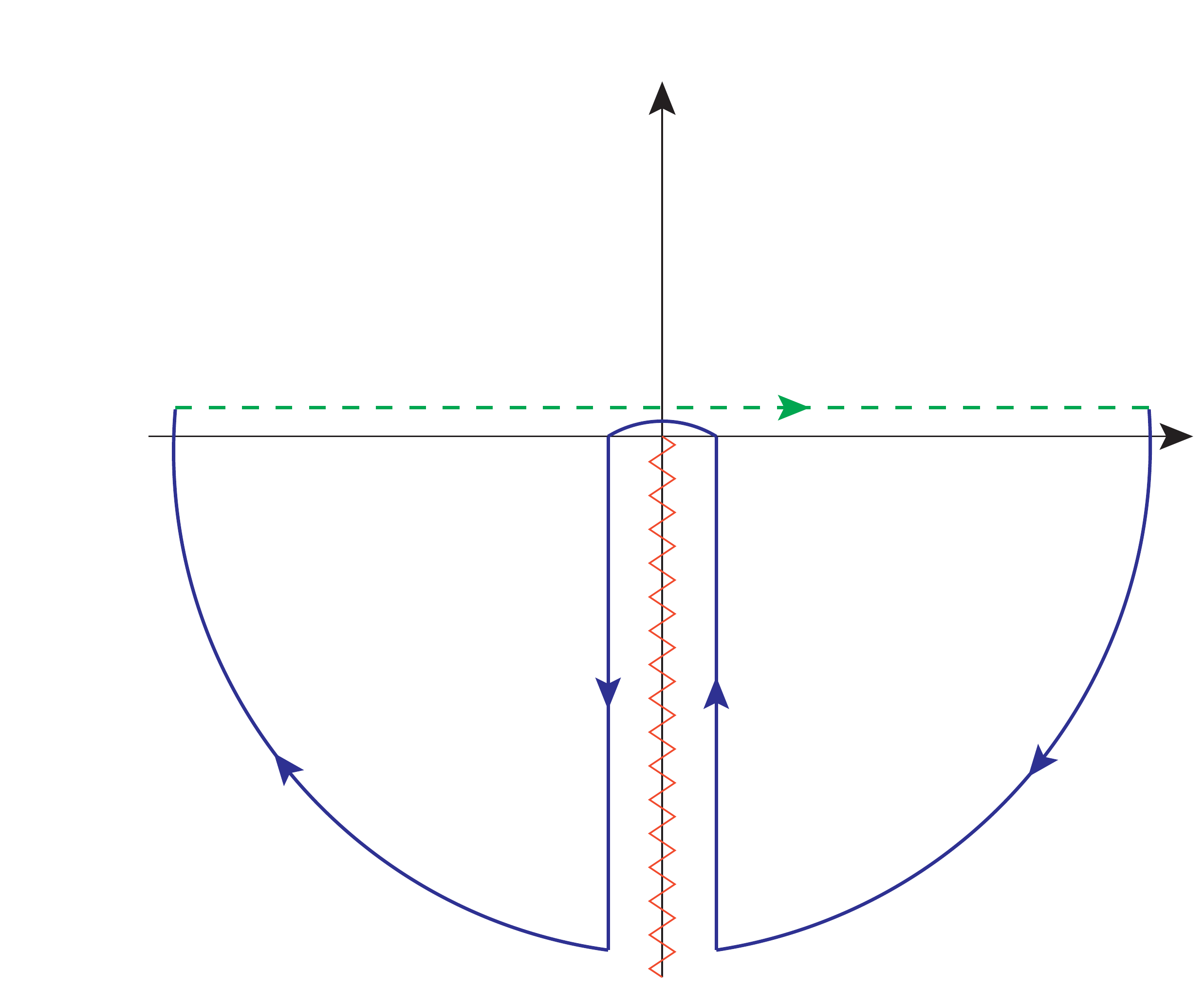}
\caption[]{Different possible contours are represented by dashed (green) and solid (blue) lines. The branch cut is represented by the zigzag (red) line.}
 \label{fig:contour}
\end{figure}

Radiation in odd spacetime dimensions - as in any dimension - is given in the frequency domain by eq. (\ref{Radiation Phi using feynman}).
However, radiation in the time domain will look essentially different, as we know that in odd $d$ an additional effect of indirect (off the lightcone) propagation comes into play.
In our analysis, this effect appears in (\ref{Radiation Phi using feynman}) via non-analyticity of the frequency domain solution for the field, in the form of a branch cut.
An inverse Fourier transform of the solution can now have two distinct contributions (see fig. \ref{fig:contour}): one from the integral along the branch cut, which gives rise to a tail term, and the other from the arc at $|\w| \! \to \! -\infty\,(\Im{\w}<0)$, which may give rise to a local term.
Fourier transforming as described gives the radiation at infinity
\bea
\Phi(\vec{r},t) = \frac{- G}{ \sqrt{2 \pi} \hd!!} r^{-\frac{\hd+1}{2}}
\sum_L n^L \d_t^{\ell+\frac{\hd}{2}} \int^{t-r}_{-\infty} \frac{Q_L(t')}{\left| t-r-t' \right|^{1/2}} \, dt' ~~  \, , ~
\label{radiation Phi odd}
\eea
which converges at both integration limits under reasonable assumptions for $Q_L(t)$.

In the case of odd spacetime dimensions (even $\hd$), $\al=\ell+\frac{\hat{d}}{2}=n$ is an integer, and (\ref{Bessel Y series2}) for $\tldya$ must be replaced with (\ref{Bessel Y series integer}).
This is of importance in the analysis of the RR effective action (\ref{S Phi using feynman},\ref{G definition}).
Multiplying the Bessel functions $\tilde{j}_n (\w r)$,~$\tilde{j}_n (\w r')\!+\!i\tilde{y}_n (\w r')$ and taking the limit $r,r'\to 0$ at the same rate\footnote{This makes sense from the point of view of the effective radiation zone theory, since the ``zoom out'' procedure out to the radiation zone affects both hatted and unhatted sources in the same manner.
From the point of view of the full theory, for a single point source and its hatted counterpart obviously $r=r'$.
For a composite source, however, there is no unique radial position of the source.
This is strongly related to the fact that in general the local term in the RR effective action depends on the source's short-distance details.},
we again regularize the finite number of terms of joint negative degree, ignore the infinite number of positive degree  terms (each giving zero), and are left only with a $\ln$ term and with constant terms,
\bea
\IGn &=& 1 + \frac{i}{\pi}
		\[ \left. 2 \ln \( \frac{\w r'}{2} \) \right|_{r' \to 0}
			- \( \psi(1)+\psi(n+1) \)
			- \sum_{m=1}^n \frac{(-)^m n!^2 }{m(n+m)!(n-m)!}
		\] \nonumber\\
&=& 1 + \frac{i}{\pi}
		\[ \left. 2 \ln \( \frac{\w r'}{2} \) \right|_{r' \to 0} +2\gm+H(2n)-2H(n) \] \, \, ,
\label{G expression odd d}
\eea
where $H(n)$ is the $n$'th harmonic number. In the frequency domain, therefore, the RR effective action is given by (\ref{S Phi using feynman}) with $\IG$ given in (\ref{G expression odd d}).
To obtain the time-domain representation of the RR effective action, we integrate over $\w$. Again, we are presented with the branch-cut discontinuity of the $\ln$ term, absent from the even dimensional case.
Placing the branch cut along the negative imaginary $w$ axis ($\w\!=\!-i\sigma$, with $\sigma \!\in\! \mathbb{R},\sigma\!\geq\!0$), we use contour integration (as shown in figure \ref{fig:contour})
to find the (dimensionally regularized) identity
\bea
\int\!\! \frac{d \w}{2 \pi} \ln (\w r') e^{-i \w (t-t')} = - \frac{\Theta(t-t')}{t-t'} +  \left( i \frac{\pi}{2} - \gamma + \ln\left( \frac{r'}{\mu} \right) \right) \delta(t-t') \, \, ,
\label{fourier of log}
\eea
where $\mu$ is some arbitrary timescale of the system.
Using the definition of the Fourier transform $Q_{L \w} = \int_{-\infty}^{+\infty} e^{i \w t} Q_{L}(t) dt$, the integral (\ref{fourier of log}) and the expressions (\ref{G expression odd d},\ref{Phi propagator scalar normalization}), from (\ref{S Phi using feynman}) we obtain \emph{the odd-dimensional radiation-reaction effective action} in the time domain,
\bea
\hS_\Phi=
G\!\!\int_{-\infty}^{\infty}\!\!\!\!\!\!\!\!{dt}\sum_{L}\! \frac{(-)^{\ell+\frac{\hd}{2}}}{\hd!!(2\ell+\hd)!!} \, \hat{Q}^L(t)
\[
	\(\frac{1}{2}H(2\ell+\hd)-H(\ell+\frac{\hd}{2}) 	+\left.\ln \left( \frac{r'}{\mu} \right)\right|_{r' \to 0} \) \d_t^{2\ell+\hd}Q_L(t) \right. \nonumber\\
\left.
	-\int_{-\infty}^{t}\!\!\!\! dt' \( \frac{1}{t-t'} \d_{t'}^{2\ell+\hd}Q_L(t') \)
\]~ . ~~~~~~~~~~~~~~~~~~~~~~~~~~
\label{S Phi scalar multipoles odd dimension}
\eea
The $\ln \left( r' / \mu \right)$ term acts as a regulator for the integral - it takes care of its divergence near the coincidence limit $t' \! \to \! t$, as the size of the system ($c=1$) is also roughly the natural cutoff for the $t'$ integration.
However, it also generically adds a finite local term of the same form as the local terms already present in (\ref{S Phi scalar multipoles odd dimension}).
The value of this additional term depends on the short-scale structure of the sources (as, for example, in \cite{FoffaSturani4PNa}).
In principle it should be determined through matching to the full theory (c.f. \cite{GoldbergerRothstein1}).
It would be interesting to start by analyzing the simple case of a single point particle in the full theory, which would also require Detweiler-Whiting (\cite{Detweiler:2002mi}) regularization.\\
One can write the RR effective action, therefore, implicitly as
\bea
\hS_\Phi=
G\!\!\int_{-\infty}^{\infty}\!\!\!\!\!\!\!\!{dt}\sum_{L}\! \frac{(-)^{\ell+\frac{\hd}{2}}}{\hd!!(2\ell+\hd)!!} \, \hat{Q}^L(t)
\[
	\(\frac{1}{2}H(2\ell+\hd)-H(\ell+\frac{\hd}{2}) \) \d_t^{2\ell+\hd}Q_L(t) \right. \nonumber\\
\left. \left.
	-\int_{-\infty}^{t}\!\!\!\! dt' \( \frac{1}{t-t'} \d_{t'}^{2\ell+\hd}Q_L(t') \) \right|_{regularized}
\]~ . ~~~~~~~~~~~~~~~~~~~~~~~~~~
\label{S Phi scalar multipoles odd dimension 1}
\eea

Thus we find that in odd spacetime dimensions the action includes both a local (purely conservative) term and a non-local term, which is responsible for dissipative effects.
The non-local (but causal!) term is a manifestation, in our effective theory of multipoles, of field propagation inside the lightcone (the so-called ``tail" effect) in the spacetime picture.
It is known that integrating out a massless degree of freedom generically leads to a non-local effective action \cite{Shuryak:2011tt,Balazs,Ching:1995tj}, as was shown to be the case in odd spacetime dimensions.
The even-dimensional case turns out to be the special case where the effective action \emph{is still local} after this elimination.

\subsection{Applications and tests}
\label{Applications and tests}

{\bf Perturbative expansion of the RR force}.
Consider the case of a single charged body with a prescribed trajectory (``being held and waved at the tip of a wand'') interacting with a scalar field.
In this case the notion of RR force and self-force coincide.
For 4d, the fully relativistic force expression is given analogously to the Abraham-Lorentz-Dirac (ALD) self force \cite{JacksonALD,Dirac} familiar from electromagnetism:
\begin{equation}
\label{F ALD 4d}
F_{ALD}^\mu \equiv \frac{d p^\mu}{d\tau} = \frac{1}{3} G\, q^2\, \( \frac{d^3 x^\mu}{d\tau^3} - \frac{d^3 x^\nu}{d\tau^3} \frac{dx_\nu}{d\tau}\,  \frac{dx^\mu}{d\tau}\).
\end{equation}
Expansion of this equation to leading and +1PN orders yields
\bea
\vec{F}_{ALD}= G\,q^2  \[  \frac{1}{3}\dot{\vec{a}}+
\frac{1}{3}v^{2}\dot{\vec{a}} + (\vec{v} \cdot \vec{a})\vec{a} + \frac{1}{3}(\vec{v} \cdot \dot{\vec{a}})\vec{v} \] .
\label{F ALD 4d scalar LO,NLO}
\eea
Upon substituting $\hd=1$, all the expressions given in sections \ref{spherical waves and double-field action} and \ref{Outgoing radiation and the RR effective action} can be seen to yield expressions identical to those given in paper I (section (3.1.2)) for the scalar field; in particular, the 4d ALD force at leading and next-to-leading order (\ref{F ALD 4d scalar LO,NLO}) is recovered immediately.\\
We derive the formula for the RR force in general even dimensions, and also check explicitly $d=6$, for which an ALD-like radiation-reaction 4-force on a scalar charge was developed by Galt'sov \cite{Galtsov:2007zz,Galtsov:2010cz}.
Our method derives the RR force from the action and multipoles (\ref{I Phi scalar multipoles},\ref{S Phi scalar multipoles}) in three stages: by using a source term of a point particle for $\rho$, by matching the appropriate $\hat{\rho}$, and by finally calculating the contribution from generalized Euler-Lagrange equation $\delta S / \delta \hat{x}^i$.
The source term corresponding to a scalar-charged point particle with a trajectory $\vec{x}(t)$ in $d$ spacetime dimensions is
\bea
\rho(\vec{x} \, ' ,t) = q \int{\delta^{(d)}(x'-x) d\tau} = \frac{q}{\gamma}\delta^{(D)}(\vec{x} \, ' - \vec{x}) = \sum_{s=0}^{\infty} \frac{-(2s-3)!! v^{2s}}{(2s)!!} \delta^{(D)}(\vec{x} \, ' - \vec{x}) \,  .
\label{scalar point source}
\eea
Thus we find
\bea
Q^L&=&-(2\ell+\hd)!!\sum_{p=0}^\infty\sum_{s=0}^\infty \frac{(2s-3)!!}
	{(2p)!!(2\ell+2p+\hd)!!(2s)!!}
	\d_t^{2p}(v^{2s}r^{2p}{x}^L_{TF}),	\nonumber\\
\hat{Q}^L&=&-(2\ell+\hd)!!\sum_{\hat{p}=0}^\infty\sum_{\hat{s}=0}^\infty \frac{(2\hat{s}-3)!!}
	{(2\hat{p})!!(2\ell+2\hat{p}+\hd)!!(2\hat{s})!!}
	\d_t^{2\hat{p}}\frac{\delta}{\delta x^{i}}(v^{2\hat{s}}r^{2\hat{p}}{x}^L_{TF})\hat{x}^{i}
\eea
Accordingly we obtain the Lagrangian, which after moving $2\hat{p}$ time derivatives from the $\hat{x}^L$ multipoles to the ${x}^L$ multipoles by partial integration becomes
\bea
\hat{L}_\Phi=G q^2	\!\sum_{L}	\!	(-)^{\ell+\frac{\hd+1}{2}}\frac{(2\ell+\hd)!!}{\hd!!}	\!
\sum_{p=0}^\infty\sum_{s=0}^\infty\sum_{\hat{p}=0}^\infty\sum_{\hat{s}=0}^\infty
	&\!\! \frac{(2s-3)!!}{(2p)!!(2\ell+2p+\hd)!!(2s)!!}
		\frac{(2\hat{s}-3)!!}{(2\hat{p})!!(2\ell+2\hat{p}+\hd)!!(2\hat{s})!!}
		\,\,\,\,\,\,\,\,\,\,\,\,\,\,\,\,\,\,\,\,\,\,\,\,\,\,	\nonumber\\
\times&\!\! \hat{x}^{i} \!\frac{\delta}{\delta x^{i}}(v^{2\hat{s}}r^{2\hat{p}}{x}^L_{TF})
		\d_t^{2(\ell+p+\hat{p})+\hd}(v^{2s}r^{2p}{x}_L)
		~.\,\,\,\,\,\,\,\,\,\,\,\,\,\,\,
\label{scalar lagrangian}
\eea
In the EOM, the RR force contribution is given by the variation by $\hat{x}^j$,
\bea
F^j=\frac{\delta \hat{S}}{\delta \hat{x}^j}=
	 \[ \frac{\d  \hat{L}}{\d  \hat{x}^j}
			- \frac{d}{dt} \( \frac{\d \hat{L}}{\d \dot{\hat{x}}^j} \)
	 \] .
\label{scalar EL}
\eea
The leading order arises from the same term (the leading dipole term $\ell=1,p=\hat{p}=s=\hat{s}=0$) in every (even) dimension, and is given by
\bea
\vec{F}^{(d)}_{LO} = \frac{(-)^{\frac{d}{2}} G q^2}{(d-1)!! (d-3)!!} \d_t^{d-1}\vec{x}~.
\label{F scalar LO any dimension}
\eea
The term for $d=6$ ($\hd=3$) is shown in table \ref{table:scalar multipoles leading}.
Out of the 15 possible action terms in the next-to-leading order (for different $\ell,p,\hat{p},s,\hat{s}$), we find using the Euler-Lagrange equation 9 non-zero contributions to the force (recorded in table \ref{table:scalar multipoles next to leading scalar} for $d=6$).
Adding these contributions we find
\bea
F_{RR}^i &=
-Gq^2& \[
\frac{1}{45} \dddot{a}^i	+
\frac{2}{45}v^2\dddot{a}^i	+
\frac{2}{9}({\vec v}\cdot {\vec a}) \ddot{a}^i	+
\frac{2}{9}({\vec v}\cdot \dot{\vec a})\dot{a}^i	+
\frac{1}{9}({\vec v}\cdot \ddot{\vec a}) a^i
\right. \nonumber\\
&&
\left. \,\,
+\frac{1}{45}({\vec v}\cdot \dddot{\vec a}){v^i}+
\frac{1}{9}({\vec a} \cdot \dot{\vec a}) a^i	+
\frac{1}{9}a^2\dot{a}^i
 \] 	.
\label{F Our 6d scalar LO,NLO}
\eea
We compared this result (\ref{F Our 6d scalar LO,NLO}) to Galt'sov's expression for the 6-dimensional scalar d-force in flat space.
We seem to find sign mismatches between his results from 2007 \cite{Galtsov:2007zz} and 2011 \cite{Galtsov:2010cz}; we assume an expression identical to his up to the signs of two subleading terms ($(\dddot{x}\ddot{x})\ddot{x}^\mu,\ddot{x}^2\dddot{x}_\nu$),
\bea
f_{flat}^\mu =
- \( \eta^{\mu\nu} + \dot{x}^\mu\dot{x}^\nu  \)   \[ \frac{1}{45} x_\nu^{(5)} - \frac{1}{9}\ddot{x}^2\dddot{x}_\nu   \]
+ \frac{2}{9} \( \dddot{x}\ddot{x} \)  \ddot{x}^\mu	,
\eea
where $\eta_{\mu\nu}$ is the flat (mostly-plus) metric  and with the derivatives taken w.r.t proper time.
Expansion of this expression up to 1-PN (next-to-leading) order to find the D-force gives
\bea
F_{Galt'sov}^i &=
-& \[
\frac{1}{45} \dddot{a}^i	+
\frac{2}{45}v^2\dddot{a}^i	+
\frac{2}{9}({\vec v}\cdot {\vec a}) \ddot{a}^i	+
\frac{2}{9}({\vec v}\cdot \dot{\vec a})\dot{a}^i	+
\frac{1}{9}({\vec v}\cdot \ddot{\vec a}) a^i
\right. \nonumber\\
&&
\left. \,\,
+\frac{1}{45}({\vec v}\cdot \dddot{\vec a}){v^i}+
\frac{1}{9}(\dot{\vec a} \cdot {\vec a}) a^i	+
\frac{1}{9}a^2\dot{a}^i
 \] ,
\label{F Galtsov 6d scalar LO,NLO}
\eea
where all derivatives are now with respect to $t$, namely $v^i:=dx^i/dt,\, a^i:=d^2x^i/dt^2$, etc.
This result coincides exactly with our result (\ref{F Our 6d scalar LO,NLO}) to +1PN order.\\
\begin{table}[h!]
  \centering \caption{Leading order contribution to the scalar self-force}
\begin{center}
\begin{tabular}{cccccccc}
  \hline
  $\ell$ $p$ $\hat{p}$ $s$ $\hat{s}$ & $\hat{L}/(G\, q^2)$ & $F^j/(G\, q^2)$ \\  \hline
  1 0 0 0 0 & $-\frac{1}{45} \hat{x}^{i}\d_t^5 x_{i}$ & $-\frac{1}{45}\dddot{a}^j$	\vspace{1mm} \\  \hline
\label{table:scalar multipoles leading}
\end{tabular}
\end{center}
\end{table}
\begin{table}[h!]
  \centering \caption{Next-to-Leading order contribution to the scalar self-force}
\begin{center}
\begin{tabular}{cccccccc}
  \hline
  $\ell$ $p$ $\hat{p}$ $s$ $\hat{s}$ & $\hat{L}/(G\, q^2)$ & $F^j/(G\, q^2)$ \\  \hline
  2 0 0 0 0 & $\frac{1}{630}\hat{x}^{k} \frac{\delta}{\delta x^{k}}[x^i x^j-\frac{1}{5} x^2 \delta^{ij}]\d_t^7(x_i x_j)$
	& $\frac{1}{315}[\d_t^7(x^i x^j)x_i-\frac{1}{5}\d_t^7(x^2)x^j]$	\vspace{1mm} \\  \hline
  1 1 0 0 0 & $-\frac{1}{630} \hat{x}^i \d_t^7 (x^2 x_i)$
	& $-\frac{1}{630} \d_t^7 (x^2 x^j)$	\vspace{1mm} \\  \hline
  1 0 1 0 0 & $-\frac{1}{630} \hat{x}^{k} \frac{\delta}{\delta x^{k}} [x^i x^2] \d_t^7 x_i$
	& $-\frac{1}{630}[x^2 \d_t^7 x^j+2x^j x_i \d_t^7 x^i]$	\vspace{1mm} \\  \hline
  1 0 0 1 0 & $+\frac{1}{90}\hat{x}^i \d_t^5(v^2 x_i)$
	& $+\frac{1}{90}\d_t^5(v^2 x^j)$	\vspace{1mm} \\  \hline
  1 0 0 0 1 & $+\frac{1}{90}\hat{x}^{k} \frac{\delta}{\delta x^{k}}[v^2 x^i] \d_t^5 x_i$
	& $+\frac{1}{90}v^2 \d_t^5 x^j -\frac{1}{45}\d_t[v^j x^i \d_t^5x_i]$	\vspace{1mm} \\  \hline
  0 1 1 0 0 & $+\frac{1}{450}\hat{x}^j x_j\d_t^7 x^2$
	& $+\frac{1}{450}x^j \d_t^7 x^2 $	\vspace{1mm} \\  \hline
  0 1 0 0 1 & $-\frac{1}{90} \hat{v}^j v_j \d_t^5 x^2$
	& $+\frac{1}{90}\d_t [v^j \d_t^5 x^2]$		\vspace{1mm} \\  \hline
  0 0 1 1 0 & $-\frac{1}{90} \hat{x}^j x_j \d_t^5  v^2$
	& $-\frac{1}{90} x^j \d_t^5 v^2 $	\vspace{1mm} \\  \hline
  0 0 0 1 1 & $+\frac{1}{18} \hat{v}^j v_j \d_t^3 v^2 $
	& $-\frac{1}{18}\d_t[v^j\d_t^3 v^2]$ 	\vspace{1mm} \\  \hline
\label{table:scalar multipoles next to leading scalar}
\end{tabular}
\end{center}
\end{table}

We can also use the multipole formulation (\ref{S Phi scalar multipoles}, \ref{I Phi scalar multipoles hat}, \ref{scalar lagrangian}, \ref{scalar EL}) to calculate the power dissipated by the RR force in even dimension $d$:
\bea
P_{RR} &=&-\vec{v}\cdot \vec{F}
=-\frac{d x^{i}}{dt} \left.\frac{\delta \hat{L}}{\delta \hat{x}^{i}}\right|_{\hat{\vec{x}} \to \vec{x}}
= G\sum_{L} \frac{(-)^{\ell+\frac{\hd-1}{2}}}{\hd!!(2\ell+\hd)!!}\frac{\delta Q^L}{\delta x^{i}}\frac{d x^{i}}{dt} \d_t^{2\ell+\hd}Q_L.
\label{radiated energy scalar definition}
\eea
The time-averaged power is found using
\bea
\int\!\!{dt} \frac{d x^{i}}{dt} \frac{\delta Q^L}{\delta x^{i}} = \int\!\!{dt} \frac{d Q^L}{dt},
\label{EL as full derivative}
\eea
followed by $\ell+\frac{\hd-1}{2}$ integrations by parts, to be
\bea
\dot{E}=\,<\!P_{RR}\!>\,  &=& \sum_{L} \frac{G}{\hd!!(2\ell+\hd)!!} \left<(\d_t^{\ell+\frac{\hd+1}{2}}Q^L )^2\right> ~,
\eea
which agrees with (\ref{radiated energy scalar multipoles}).
In 6d, we record the radiated power to +1PN order,
\bea
\dot{E}=Gq^2 \[ \frac{1}{45}\dot{a}^2
	+ \frac{1}{15}v^2 ({\vec v} \!\cdot\! \dddot{\vec a})
	+ \frac{1}{3}({\vec v} \!\cdot\! {\vec a})({\vec v} \!\cdot\! \ddot{\vec a})
	+ \frac{1}{9}({\vec a} \!\cdot\! \dot{\vec a})({\vec v} \!\cdot\! {\vec a})
	+ \frac{1}{9}a^2 ({\vec v} \!\cdot\! \dot{\vec a})
	+ \frac{2}{9}({\vec v} \!\cdot\! \dot{\vec a})^2  \] \!.~~~~~
\label{P Our 6d scalar LO,NLO}
\eea
We note the special case of $2d$, where the only STF multipoles are $\{1,x\}$; the leading order remains (\ref{F scalar LO any dimension}), while at higher orders the sum over $L$ trivializes.

\section{Electromagnetism}
\label{section:Electromagnetism}
The EM action is given by
\bea
\label{MaxwellEMAction}
S= -\frac{1}{4 \Omd} \int\!\! F_{\mu \nu} F^{\mu \nu} r^{\hat{d}+1} dr \dOmd dt  - \int\!\! A_{\mu} J^{\mu}  r^{\hat{d}+1} dr \dOmd dt\, .
\eea
Working in spherical coordinates $(t,r,\Om)$, we reduce over the sphere as in (\ref{decomposition of scalar field},\ref{decomposition of scalar field and source}). The EM field and sources are decomposed as
\bea
A_{t/r}&=&\int\frac{d \w}{2\pi}\sum_L A_{t/r}^{L \w}  x_L e^{-i \w t} \, , \nonumber \\
A_{\Om}&=&\int\frac{d \w}{2\pi}\sum_L  \(  A_{S}^{L \w} \partial_{\Om} x_L + A_\Valph^{L \w} x^{L}_\AlOm  \)   e^{-i \w t} \, , \nonumber \\
J^{t/r}&=&\int\frac{d \w}{2\pi}\sum_{L} J^{t/r}_{L \w}  x^L e^{-i \w t} \, , \nonumber \\
J^{\Om}&=&\int\frac{d \w}{2\pi}\sum_L  \(  J^{S}_{L \w} \partial^{\Om} x^L + J^\Valph_{L \w} x_{L}^{\AlOm}  \)   e^{-i \w t} \, ,
\label{decomposition of EM field and sources}
\eea
where the scalar multipoles $x^L$ (\ref{scalar spherical harmonics normalization},\ref{scalar spherical harmonics normalization 2}) are now supplemented by the divergenceless vector multipoles $x^{L}_\AlOm$, enumerated by an antisymmetric multi-index $\aleph$ taken from the Hebrew alphabet, representing $D-3$ spherical indices:
\bea
x^{L}_\AlOm=\eps^{(\hd+1)}_{\AlOm \, \Om'} \, \d^{\Om'} \, x^L=\eps^{(D)}_{\AlOm \, b \, c}x^b \, \d^c \, x^L=\delta^a_\Om \, \eps^{(D)}_{\aleph \, a \, b \, c} \, x^b \, \d^c \, x^L=\( \star (\vec{r} \wedge \vec\nabla )\)_{\!\AlOm}  \!\!\!\!x^L	~,
\label{x L aleph Omega}
\eea
where $\eps^{(\hd+1)}_{\Om_1\cdots\Om_{\hd+1}}$ ($\eps^{(D)}_{a_1\cdots a_D}$) is the complete antisymmetric tensor on the $\Omd$-sphere (in D spatial dimensions), $\wedge$ is the exterior product and $\star$ is the Hodge duality operator \cite{{Hestenes1,Hestenes2,Baylis,Doran,Hestenes3,Rowland,Chappell}}.
The dimension of this independent vectorial basis is $D_\ell(\hd+1,1) = \frac{\ell(\ell+\hd)(2\ell+\hd)(\ell+\hd-2)!}{(\hd-1)!(\ell+1)!}$ \cite{RubinOrdonez,Higuchi:1986wu}\footnote{These are generalizations of the single 4d ($\hd=1$) multipole family $x^{L}_\Om\!=\!\eps_{\Om \Om'}\d^{\Om'} x^L\!=\!(\vec{r}\! \, \times\!\! \, \vec\nabla x^L)_\Om$.
In 4d, $D_\ell(\hd+1,1)=2\ell+1=D_\ell(\hd+1,0)$, and thus these indeed form a single family; $\aleph$ is then an empty string.}.
The complete normalization conditions in d dimensions are \cite{RubinOrdonez,Higuchi:1986wu}
\bea
&\int& x_{L_\ell} x^{L'_{\ell '}} \dOmd
= \Nld r^{2\ell} \Omd \delta_{\ell \ell'} \delta_{L_\ell L'_{\ell'}} \, , \nonumber \\
&\int& g^{\Om \Om'} \partial_{\Om} x_{L_\ell} \partial_{\Om'} x^{L'_{\ell '}} \dOmd
= c_s \cdot \Nld r^{2\ell} \Omd \delta_{\ell \ell'} \delta_{L_\ell L'_{\ell'}} \, , \nonumber \\
&\int& g^{\Om \Om'} x^{L_\ell}_{\aleph \Om} x^{L'_{\ell'}}_{\aleph' \Om'} \dOmd
= c_s \cdot \Nld r^{2\ell} \Omd \delta_{\ell \ell'} \delta_{L_\ell L'_{\ell'}} \delta_{\aleph\aleph'} \, ,\nonumber \\
&\int&  g^{\Om \Om'} g^{\Psi \Psi'} x^{L_\ell}_{\aleph [\Psi;\Om]} x^{L'_{\ell'}}_{\aleph' [\Psi';\Om']} \dOmd
= 2 c_v c_s \cdot \Nld \Omd r^{2\ell} \delta_{\ell \ell'} \delta_{L_\ell L'_{\ell'}} \delta_{\aleph\aleph'} \, ,
\label{vector sphercal harmonics normalization}
\eea
where $c_s:=\ell(\ell+\hd)$, $c_v=c_s+\hd-1$ and the semicolon $``;"$ represents the covariant derivative on the sphere.
The inverse transformations are given by\\
\bea
J^t_{L\w}(r)  &=&
 \(  \Nld r^{2\ell} \Omd  \) ^{\!\!-1}\!\!\!\!\int\!\! \rho_\w(\vec{r}) x_L \dOmd=
 \(  \Nld r^{2\ell} \Omd  \) ^{\!\!-1}\!\!\!\!\int\!\!\!\!\int\!\!{dt}e^{i \w t} \rho(\vec{r},t) x_L \dOmd \, ,	\nonumber\\
J^r_{L\w}(r) &=&
 \(  \Nld r^{2\ell} \Omd  \) ^{\!\!-1}\!\!\!\!\int\!\! \vec{J}_w(\vec{r})\cdot \vec{n} x_L \dOmd=
 \(  \Nld r^{2\ell} \Omd  \) ^{\!\!-1}\!\!\!\!
\int\!\!\!\!\int\!\!{dt}e^{i\w t} \vec{J}(\vec{r},t)\cdot \vec{n} x_L \dOmd,\nonumber\\
J^\Valph_{L\w}(r) &=&
 \(c_s  \Nld r^{2\ell} \Omd  \) ^{\!\!-1}\!\!\!\!\int\!\! \vec{J}_w(\vec{r})\!\cdot\!
\( \star (\vec{r} \wedge \vec\nabla )\)_{\!\aleph} \!\!  x^L \dOmd		\nonumber\\
&=& \(c_s  \Nld r^{2\ell} \Omd  \) ^{\!\!-1}\!\!\!\!
\int\!\!\!\!\int\!\!{dt}e^{i\w t} \vec{J}(\vec{r},t) \!\cdot\!  \( \star (\vec{r} \wedge \vec\nabla )\)_{\!\aleph} \!\!  x^L  \dOmd.\,\,\,\,\,\,\,\,\,\,\,\,\,\,\,
\label{EM inverse sources}
\eea
We note that only three inverse transformations are required, as we henceforth replace $J^S_{L \w}$ using current conservation
\bea
0=D_\mu J^\mu_{L\w} = i \w J^t_{L\w} + (\d_r+\frac{\ell+\hd+1}{r})J^r_{L\w}-c_s J^S_{L\w}.
\label{current conservation}
\eea
We plug (\ref{decomposition of EM field and sources}) into the action (\ref{MaxwellEMAction}) to obtain
\bea
S=\frac{1}{2}\int\frac{d \w}{2\pi}\sum_L
	 &\!\!\!\!\!\!\!\!\!\!\!\!\!\!\!\!\!\!\!\!\!\!\!\!\!\!\!\!\!\!\!\!\!\!\!\!\!\!\!\!\!\!\!\!\!\!\!\!\!\!\!\!\!\!\!\!\!\!\!\!\!\!\!\!\!\!\!\!\!\!\!\!\!\!\!\!\!\!\!\!\!\!\!\!\!\!\!\!\!\!\!\!\!\!\!\!\!\!\!\!\!\!\!\!\!\!\!\!\!\!\!\!\!\!\!\!\!\!\!\!\!\!\!\!\!\!\!\!\!\!\!\!\!\!\!\!\!\!\!\!\!\!\!\!\!\!\!\!\!\!\!\!\!\!\!\!\!\!\!\!\!\!\!\!
	 \Nld S_{L\w},	\nonumber\\
S_{L\w}=\!\int\!\!dr r^{2\ell+\hd+1} &
	\!\!\left\{
	 \[ \left| i \w A^{L\w}_r - \frac{1}{r^\ell}(r^\ell A^{L\w}_t)' \right|^2
	+\frac{c_s}{r^2}  \left| i \w A^{L\w}_S - A^{L\w}_t \right|^2
	-\frac{c_s}{r^2} \left| \frac{1}{r^\ell}(r^\ell A^{L\w}_S)' - A^{L\w}_r \right|^2 \right. \right.
\nonumber \\
	 &\!\!\!\!\!\!\!\!\!\!\!\!\!\!\!\!\!\!\!\!\!\!\!\!\!\!\!\!\!\!\!\!\!\!\!\!\!\!\!\!\!\!\!\!\!\!\!\!\!\!\!\!\!\!\!\!\!\!\!\!\!\!\!\!\!\!\!
\left.\left.+c_s \( \frac{\w^2}{r^2}-\frac{c_v}{r^4} \) \left| A^{L\w}_\Valph \right|^2
	- \frac{c_s}{r^2} \left| \frac{1}{r^\ell}(r^\ell A^{L\w}_\Valph)' \right|^2    \right.  \]
\nonumber \\
	&\!\!\!\!\!\!\!\!\!\!\!\!\!\!\!\!\!\!\!\!\!
- \left. \Omd  \[  A^{L\w}_r J_{L\w}^{r *} + A^{L\w}_t J_{L\w}^{t *}
						+ c_s A^{L\w}_S J_{L\w}^{S *} + c_s A^{L\w}_\Valph J_{L\w}^{\Valph *} + c.c.  \]  \right\}
,\!\!\!
\label{EM action spherical}
\eea
where $':=\frac{d}{dr}$, and we use $A_{L  -\w} = A^{*}_{L \w}$, $J_{L -\w} = J^{*}_{L \w}$ since $A_{\mu}(x)$, $J^{\mu}(x)$ are real.
In the spirit of \cite{AsninKol}, we notice that $A^{L\w}_r$ is an auxiliary field, a field whose derivative $A'_r$ does not appear in (\ref{EM action spherical}).
Therefore, its EOM is algebraic and is solved to yield
\bea
A_{L\w}^r = -\frac{1}{\w^2 - \frac{c_s}{r^2}}  \[ \frac{i \w}{r^\ell}(r^\ell A_{L\w}^{t})' + \frac{c_s}{r^{\ell+2}}(r^\ell A_{L\w}^{S})' - \Omd J^r_{L\w}  \]  ~.
\label{Ar}
\eea
Substituting the solution into the action, it is seen that the action can be separated into independent fields, corresponding to the different possible polarizations.
We distinguish between the vectorial $A^{L\w}_{V\aleph}$ fields, as they appear already in (\ref{EM action spherical}), coupled to the vector source terms
\bea
\rho^\Valph_{L\w} := J^\Valph_{L\w} \, \, ,
\label{EM source vector}
\eea
and the scalar fields
\bea
\tilde{A}^{L\w}_S:=A^{L\w}_t - i\w A^{L\w}_S\, ,
\label{astilde def}
\eea
which are coupled to the corresponding source terms
\bea
\rho^S_{L\w} := - J^t_{L\w}+\frac{i}{\w r^{\ell+\hd+1}}  \(  r^{\ell+\hd+1} \frac{\Lambda}{\Lambda-1} J^r_{L\w}  \) ' \, ,
\label{EM source scalar}
\eea
where $\Lambda:=\frac{\w^2 r^2}{c_s}$, and we have used (\ref{current conservation}).
The action can now be concisely decoupled to scalar and vector parts (omitting hereafter the indices $(L \w)$ for brevity):
\bea
S_{EM} = \frac{1}{2}\int\frac{d \w}{2\pi}\sum_L \[  S^{L\w}_S + \sum_\aleph
S^{L\w}_\Valph  \] ,
\label{SEM}
\eea
with
\bea
S^{L\w}_S &=& \!\! \Nld \! \int\!\!{r^{2\ell+\hd+1} dr}
 \[ \frac{1}{1-\Lambda} \left| \frac{1}{r^\ell} (r^\ell \tilde{A}_S)' \right|^2 +\frac{c_s}{r^2} \left| \tilde{A}_S \right|^2 +\Omd (\tilde{A}_S \rho^{S*}_{L\w} +c.c.)  \] ,
\label{SS}
\\
S^{L\w}_\Valph &=& \!\! \Nld \! \int\!\!{r^{2\ell+\hd+1} dr} c_s \!
 \[  \(  \frac{\w^2}{r^2} \!-\! \frac{c_v}{r^4}  \) \! \left| A_\Valph \right|^2 \!\!-\!\left| \frac{1}{r^{\ell+1}} (r^\ell \tilde{A}_\Valph)' \right|^2 \!\!\!-\!\Omd (\tilde{A}_\Valph \rho^{\Valph*}_{L\w} \!+\! c.c.)  \] \!\!,~~~~~~~
\label{SV}
\eea
and we treat them separately.
Note that for $\ell=0$ we have $\rho^S_{L\w}=0$ (see eq. (\ref{current conservation}),(\ref{EM source scalar})) as well as $S_V=0$, thus only $\ell\geq1$ need be considered.

\subsection*{The scalar part of the EM action}
\label{subsection:EM scalar}

We derive the equation of motion for the scalar action from $S^{L\w}_S$ (\ref{SS}) by treating $(r^\ell\tilde{A}_S)$ as the field, and finding equations for its conjugate momentum $ \( \Nld r^\ell \Pi_S \) $,
\bea
 \( \Nld r^\ell \Pi_S \) :=\frac{\partial L}{\partial(r^\ell \tilde{A}_S^{*})'}
&=&\frac{\Nld}{1-\Lambda}r^{\hd+1} (r^\ell\tilde{A}_S)', \label{PI S1}\\
 \( \Nld r^\ell\Pi_S \) ':=\frac{\partial L}{\partial(r^\ell\tilde{A}_S^{*})}
&=&\Nld c_s r^{\hd-1}(r^\ell\tilde{A}_S)+\Nld \Omd r^{\ell+\hd+1}\rho^S_{L\w}.
\label{PI S2}
\eea
Differentiating (\ref{PI S2}) with respect to $r$, substituting (\ref{PI S1}), and renaming the field $\PhiS$ and source term $\rho^{\PhiS}_{L\w}$ (recalling (\ref{EM inverse sources}),(\ref{EM source scalar})) as
\bea
\PhiS&=&  \( \ell \, r^\hd \) ^{-1} {\Pi_S}, \nonumber\\
\rho^{\PhiS}_{L\w}&=&\!\Nld \Omd \frac{r^{\ell+\hd}(r^{\ell+2} \rho^S_{L\w})'}{\ell+\hd}
\!=\!\!\int\!\! d\Omd \frac{r^\hd x_L}{\ell+\hd} \!  \[
\frac{i}{\w \, r^{\hd-1}}\!\!
 \(  \! r^{\hd+1} \frac{\Lambda}{\Lambda-1} \vec{J}_w(\vec{r})\!\cdot\! \vec{n} \!  \) '
\!\!\!-\! r^2 \rho_\w(\vec{r})
 \] ' \!\!\!\!,~~~~~~~ \nonumber \\
\label{EM source scalar Phi}
\eea
we find the equation
\bea
0=\Nld \, r^{2\ell+\hd+1} \frac{\ell}{\ell+\hd}	 \( \w^2+\d_r^2+\frac{2\ell+\hd+1}{r}\d_r  \)  {\PhiS}-\rho^{\PhiS}_{L \w}.
\label{EOM PhiS}
\eea
This equation is of the same form as (\ref{EOM Phi}), up to the replacement of $G$ by 
\bea
\REone&=&\frac{\ell+\hd}{\ell},
\label{REone}
\eea
thus we find a propagator similar to (\ref{Phi propagator scalar}),
\bea
G^{\PhiS}_{ret}(r',r)= \REone \, G^\Phi_{ret}(r',r)
&=&-i \w^{2\ell+\hd} \Mld \, \REone \, \tilde{j}_{\alpha}(\w r_1)\, \tilde{h}^+_{\alpha}(\w r_2) \delta_{LL'}
\, \, \, \, ; \nonumber\\
r_1:&=&\text{min}\{r',r\},\,\,\,r_2:=\text{max}\{r',r\}.
\label{EM propagator scalar}
\eea
We again present the EFT Feynman rules following the steps (\ref{scalar wavefunction at radiation zone1},\ref{scalar wavefunction at radiation zone2},\ref{I Phi scalar multipoles}).
In the radiation zone, the field can be written as
\bea
{\PhiS^{L \w}}^{EFT}(r)=  \( -Q^{E}_{L \w}  \)    \(  -i \w^{2\ell+\hd} \Mld \,\REone \tilde{h}^+_{\ell}(\w r)  \)  \, \, ,
\label{EM scalar wavefunction at radiation zone1}
\eea
where $Q^{E}_{L \w}$ are the sources (fig. \ref{vertex_definition}).
In the full theory the solution outside the sources is given (see \ref{EOM PhiS}) by
\bea
\PhiS^{L \w}(r) = \!\!\!\int\!\! dr' \rho^{\PhiS}_{L\w}(r') G^{\PhiS}_{ret}(r',r)
	=  \[  -\!\!\!\int\!\! dr' \tilde{j}_{\alpha}(\w r') \rho^{\PhiS}_{L\w}(r')  \]
		 \[ -i \w^{2\ell+\hd} \Mld \, \REone \tilde{h}^+_{\ell}(\w r) \] \!\!, \,\,\,\,\,\,\,\,\,\,\,\,
\label{EM scalar wavefunction at radiation zone2}
\eea
and the sources can be read off and identified (using (\ref{EM inverse sources}, \ref{EM source scalar Phi}), integrations by parts and \ref{Modified Bessel equation}) to be
\bea
Q^{(E)}_{L \w}&=&\int\!\!{dr'} \tilde{j}_{\alpha}(\w r') \rho^{\PhiS}_{L\w}(r')		\nonumber\\
&=&\frac{1}{\ell+\hd}\int\!\!{dr'} \tilde{j}_{\alpha}(\w r') \!\!\int\!\! d\Omd' \, r'^\hd x'_L \!\!
	 \[ - r'^2 \rho_{\w}(\vec{x} \, ')
	+\frac{i}{\w \, \, r'^{\hd-1}}  \(  r'^{\hd+1} \frac{\Lambda}{\Lambda-1} \vec{J}_\w(\vec{x} \, ')\cdot \vec{n} \, '  \) '
	 \] '	\nonumber\\
&=& \frac{1}{\ell+\hd}
	\!\int\!\!{d^D x'} x'_L
	 \[ \frac{1}{r'^{\ell+\hd-1}} \( r'^{\ell+\hd}\tilde{j}_{\alpha}(\w r') \) ' \rho_\w(\vec{x} \, ')
	-i\w \tilde{j}_{\alpha}(\w r') \vec{J}_\w(\vec{x} \, ') \cdot \vec{x} \, '
	 \] .
\eea
We return to the time domain using (\ref{EM inverse sources}) to find \emph{the electric type radiation source multipoles} (compare (\ref{I Phi scalar multipoles}))
\bea
Q^L_{(E)}&=&\frac{1}{\ell+\hd}\int\!\!{d^D x} x^L_{TF}
	 \[ \frac{1}{r^{\ell+\hd-1}} \( r^{\ell+\hd}\tilde{j}_{\alpha}(ir \d_t) \) ' \rho(\vec{x})
	-\tilde{j}_{\alpha}(ir \d_t)\, \d_t \vec{J}(\vec{x}) \cdot \vec{x} \,  \] ,
\label{I EM scalar multipoles}
\\
\hat{Q}_{(E)}^L&=&\frac{\delta Q_E^L}{\delta J^{i}}\hat{J}^{i}=\frac{\delta Q_E^L}{\delta x^{i}}\hat{x}^{i}.
\label{I EM scalar multipoles hat}
\eea
We note here that an expansion of $\tilde{j}_\alpha$ according to (\ref{Bessel J series2}) reproduces, for $d=4$, eq.(47) of \cite{RossMultipoles} (after using current conservation, eq. (49) there).

\subsection*{The vector part of the EM action}
\label{subsection:EM vector}

For the vector sector of the action, we rewrite (\ref{SV}) in a form similar to (\ref{scalar action spherical}),
\bea
S^{L\w}_\Valph = \int \!\! dr \!
 \[ \frac{\Nld \, r^{2\ell+\hd+1}}{\RMone} \PhiV^* \!  \( \! \w^2+\d_r^2+\frac{2\ell+\hd+1}{r}\d_r  \! \)  \!\! \PhiV
\!-\! \( \rho^{\PhiV}_{L\w} \PhiV^* \!+\!c.c. \)   \] \!\! ,\,\,\,\,\,\,\,\,\,\,\,
\label{S_V in Phi}
\eea
where we have defined (recalling (\ref{EM inverse sources})) as
\bea
\RMone&=&\frac{\ell}{(\ell+\hd)}~,		\label{RMone}	\\
\PhiV&=&\frac{\ell A_\Valph}{r}~,			\nonumber\\
\rho^{\PhiV}_{L\w}&=&(\ell+\hd) \Nld\,\Omd r^{2\ell+\hd+2} \rho^V_{L\w}(r)=
  \frac{1}{\ell} \, r^{\hd+2} \!\!\int\!\! \vec{J}_w(\vec{r})\!\cdot\!
\( \star (\vec{r} \wedge \vec\nabla )\)_{\!\aleph} \!\!  x_L \dOmd
~.~~~
\label{EM source vector Phi}
\eea
The action (\ref{S_V in Phi}) is again identical to (\ref{scalar action spherical}) up to a prefactor of $\RMone$, with a source similar to (\ref{scalar action source}).
The propagator, therefore, is (compare \ref{Phi propagator scalar}, \ref{EM propagator scalar})
\bea
G^{\PhiV}_{ret}(r',r)
=-i\w^{2\ell+\hd} \, \Mld \, \RMone \, \tilde{j}_{\alpha}(\w r_1) \, \tilde{h}^+_{\alpha}(\w r_2) \, \delta_{LL'}~;\nonumber\\
r_1:=\text{min}\{r',r\},\,\,\,r_2:=\text{max}\{r',r\}.
\label{EM propagator vector}
\eea
We find these sources $Q^{(M,\aleph)}_{L \w}$ by again matching $\Phi^{\PhiV}_{L \w}(r)$ for large $r$ and from the diagrammatic representation (in analogy with \ref{scalar wavefunction at radiation zone1},\ref{scalar wavefunction at radiation zone2},\ref{I Phi scalar multipoles},\ref{EM scalar wavefunction at radiation zone1},\ref{EM scalar wavefunction at radiation zone2},\ref{I EM scalar multipoles}), to find
\bea
Q^{(M,\aleph)}_{L \w}&=&\int\!\!{d^D x} \tilde{j}_{\alpha}(\w r)  \(\star(\vec{r}\!\wedge\! \vec{J}_w(\vec{r})) \)^\aleph \!\!\!\!\!_{(k_\ell} x_{L-1)},
\eea
where we have used (\ref{EM inverse sources},\ref{EM source vector}).
In the time domain, we find \emph{the magnetic radiation source multipoles} (compare (\ref{I Phi scalar multipoles}, \ref{I EM scalar multipoles})),
\bea
Q^L_{(M,\aleph)}&=&\int\!\!{d^D x} \tilde{j}_{\alpha}(ir\d_t)  \[ \( \star (\vec{r}\wedge \vec{J}(\vec{r}) ) \)_\aleph \!\!\!\!\!^{k_\ell} x^{L-1} \] ^{STF},
\label{J EM vector multipoles}
\\
\hat{Q}_{(M,\aleph)}^L&=&\frac{\delta Q_{(M,\aleph)}^L}{\delta J^{i}}\hat{J}^{i}=\frac{\delta Q_{(M,\aleph)}^L}{\delta x^{i}}\hat{x}^{i}.
\label{J EM vector multipoles hat}
\eea
Presenting $\tilde{j}_\alpha$ as a series expansion using (\ref{Bessel J series2}), these coincide with eq.(48) of \cite{RossMultipoles}.

\subsection{Outgoing EM radiation and the RR effective action}
Outgoing EM radiation can now be found diagrammatically (compare \ref{Radiation Phi using feynman}) as
\begin{align}
\label{Radiation EM using feynman scalar}
\PhiS^{L\w}(r)=&
\parbox{20mm} {\includegraphics[scale=0.5]{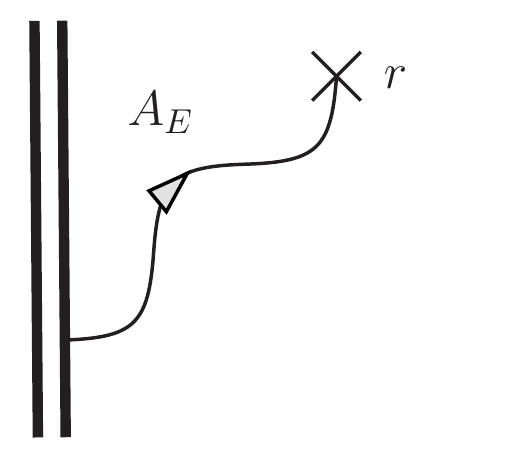}}
= 	-Q_{(E)}^{L'\w}G^{\PhiS}_{ret}(0,r)
=\sqrt{\frac{\pi}{2^{\hd+1}}} \frac{\REone}{\Gm(1+\hd/2)} (-i\w)^{\ell+\frac{\hd-1}{2}} \frac{Q_{(E)}^{L\w}}{r^{\ell}} \frac{e^{i\w r}}{r^{\frac{\hd+1}{2}}}\,\, , \,\,\,\,\,\,
\nonumber\\ 
\PhiV^{L\w}(r)=&
\parbox{20mm} {\includegraphics[scale=0.5]{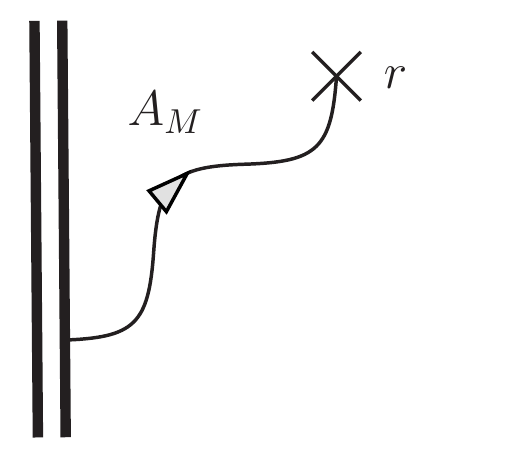}}
= 	-Q_{(M,\aleph)}^{L'\w}G^{\PhiV}_{ret}(0,r)
=\sqrt{\frac{\pi}{2^{\hd+1}}} \frac{\RMone}{\Gm(1+\hd/2)} (-i\w)^{\ell+\frac{\hd-1}{2}} \frac{Q_{(M,\aleph)}^{L\w}}{r^{\ell}} \frac{e^{i\w r}}{r^{\frac{\hd+1}{2}}}\,\, . \,\,\,\,\,\,
\end{align}
In the time domain, for even $d$, we find (where $\eps$ is $(+)/(- , \aleph)$ for the electric/magnetic part/s)
\bea
A_\eps(\vec{r},t) =\frac{1}{\hd!!} r^{-\frac{\hd+1}{2}}
\sum_L R^\eps_1 n^L \d_t^{\ell+\frac{\hd-1}{2}} Q^\eps_L(t-r) ~ \, . ~
\label{radiation A}
\eea
The EM double field effective action is a sum of the scalar and vector action diagrams and can be written using our Feynman rules, similarly to (\ref{S Phi using feynman},\ref{G expression non-odd d}),
\begin{align}
\label{S EM multipoles}
\hS_{EM}=&\,\,\,\,\,\,\,\,\,\,\,\,\,\,\,\,\,\,\,\,\,\,\,\,\,\,\,\,\,\,\,\,\,\,\,\,\,\,\,\,\,\,
\parbox{20mm} {\includegraphics[scale=0.5]{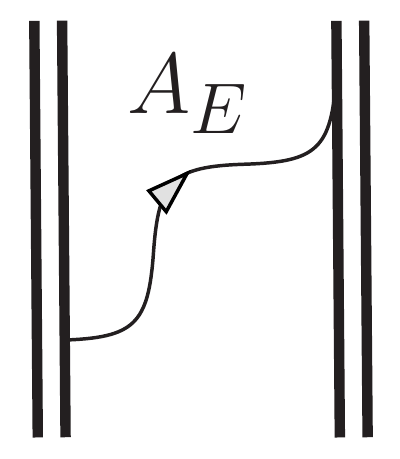}}
\,\,\,\,\,\,\,\,\,\,\,\,\,\,\,\,\,\,\,\,\,\,\,\,\,\,\,\,\,\,\,\,\,+\,\,\,\,\,\,\,\,\,\,\,\,\,\,\,\,\,\,\,\,\,\,\,\,\,\,\,\,
\parbox{20mm} {\includegraphics[scale=0.5]{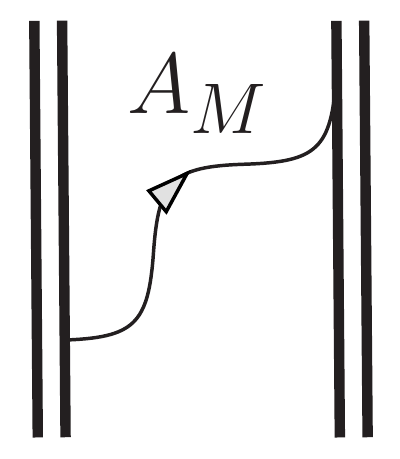}}
\nonumber\\
=& \frac{1}{2}\int\!\frac{d \w}{2\pi}\sum_{L,L'}
	 \[
		 \( -Q^{(E)}_{L\w} \)
		G^{\PhiS}_{ret}(0,0)
		 \( -\hat{Q}^{(E)*}_{L'\w} \)
	\!\!+\!\!
		 \( -Q^{(M,\aleph)}_{L\w} \)
		G^{\PhiV}_{ret}(0,0)
		 \( -\hat{Q}^{(M,\aleph)*}_{L'\w} \)
	 \]
	\!\!+\!\!c.c.\,\,\,\,\,\,\,\,\,\,
	\nonumber\\
=& \!\int\!\frac{d \w}{2\pi}\sum_L
	\frac{-i \w^{2\ell+\hd} \Mld \cdot\IG(\w) }{2}
	 \[
		\REone	Q^{L\w}_{(E)} \hat{Q}^{(E)*}_{L\w}
	+
		\RMone 	Q^{L\w}_{(M\aleph)} \hat{Q}^{(M\aleph)*}_{L\w}
	 \]
	+c.c.~,
\end{align}
where $Q^L_{(E)},\hat{Q}^L_{(E)},Q^L_{(M)},\hat{Q}^L_{(M)}$ were given by (\ref{I EM scalar multipoles},\ref{I EM scalar multipoles hat},\ref{J EM vector multipoles},\ref{J EM vector multipoles hat}).
In even $d$, we can transform to the time domain and use (\ref{Phi propagator scalar normalization},\ref{REone},\ref{RMone},\ref{G expression non-odd d}) explicitly to find
\bea
\hS_{EM}=\int\!\!{dt}\!\sum_{L}\frac{(-)^{\ell+\frac{\hd+1}{2}}}{\hd!! (2\ell+\hd)!!}
	 \[
		\frac{\ell+\hd}{\ell} \hat{Q}^{(E)}_L \cdot \d_t^{2\ell+\hd}Q_{(E)}^L
	+
		\frac{\ell^2 \hd}{(\ell\!+\!1)(\ell\!+\!\hd\!-\!1)} \hat{Q}^{(M)}_L \cdot \d_t^{2\ell+\hd}Q_{(M)}^L
	 \] \!\! ,~~~~~
\label{S EM multipoles time}
\eea
where we have also summed over the $\aleph$ indices, defining the bi-vector multipoles
\bea
Q^L_{(M)} = \int\!\!{d^D x} \tilde{j}_{\alpha}(i r \d_t)  \( \vec{r}\!\wedge\! \vec{J}(\vec{r}) \) \!^{(k_\ell} x^{L-1)},
\eea
defining $\hat{Q}^L_{(M)}$ accordingly, and incurring a factor of
\bea
\frac{D_\ell (\hd+1,1)}{D_\ell (\hd+1,0)}=\frac{\ell\hd(\ell+\hd)}{(\ell+1)(\ell+\hd-1)}
\eea
from summing over the different $\aleph$ combinations.\\
The case of the EM field in non-even (and in particular odd) spacetime dimensions is treated in a similar manner to the scalar case (Sec.\ref{Odd spacetime dimensions}), and similar non-local ``tail" expressions appear (compare with (\ref{S Phi scalar multipoles odd dimension 1})):
\bea
\hS_{EM}&=&\int\!\!{dt}\!\sum_{L}\frac{(-)^{\ell+\frac{\hd+1}{2}}}{\hd!! (2\ell+\hd)!!}
	 \[
		\frac{\ell+\hd}{\ell} S^{(E)}(t)
	+
		\frac{\ell^2 \hd}{(\ell\!+\!1)(\ell\!+\!\hd\!-\!1)} S^{(M)}(t)
	 \] \!\! ,~~~~
\label{S EM multipoles odd dimension 1}
\\  \nonumber\\ 
S^{(E/M)}(t) &=& \hat{Q}^{(E/M)}_L(t)
\[
	\(\frac{1}{2}H(2\ell+\hd)-H(\ell+\frac{\hd}{2}) \) \d_t^{2\ell+\hd}Q_{(E/M)}^L(t)  \right. \nonumber\\ 
&& \left. \left. ~~~~~~~~~~~~~~
	-\int_{-\infty}^{t}\!\!\!\! dt' \( \frac{1}{t-t'} \d_{t'}^{2\ell+\hd}Q_{(E/M)}^L(t') \) \right|_{regularized}
\]~ . ~
\label{S EM multipoles odd dimension 2}
\eea

\subsection{Applications and tests}	

{\bf Perturbative expansion of the RR force and comparison with ALD}.
For the RR force on a single accelerating electric charge we have the ALD formula \cite{Dirac} in $4d$,
\begin{equation}
\label{EM ALD}
F_{ALD}^\mu \equiv \frac{d p^\mu}{d\tau} = \frac{2}{3} q^2\, \( \frac{d^3 x^\mu}{d\tau^3} - \frac{d^3 x^\nu}{d\tau^3} \frac{dx_\nu}{d\tau}\,  \frac{dx^\mu}{d\tau}\).
\end{equation}
Our expressions, specialized to $d=4$, can be seen to be identical to those given in paper I (section (3.2.2)), already shown to reproduce the ALD result.
We test here for $d=6$, comparing the force with Galt'sov's result \cite{Galtsov:2007zz}:
\bea
f_{flat}^\mu &=&
- \( \eta^{\mu\nu} + \dot{x}^\mu\dot{x}^\nu  \)
	 \( \frac{4}{45} x_\nu^{(5)} - \frac{2}{9}\ddot{x}^2\dddot{x}_\nu  \)
+\frac{2}{3} \( \dddot{x}\ddot{x} \)  \ddot{x}^\mu~.
\label{F Galtsov 6d EM full}
\eea
Expanded to leading and next-to-leading order, we find
\bea
F^i_{Galt'sov} &=&
- \[
	\frac{4}{45}{a^i}'''
	+\frac{8}{45}v^2{a^i}'''
	+\frac{8}{9}({\vec v}\!\cdot\! {\vec a}) {a^i}''
	+\frac{8}{9}({\vec a}'\!\cdot\! {\vec v}){a^i}'
\right.\nonumber\\ && ~~~~\left.
	+\frac{4}{45}({\vec a}'''\!\cdot\! {\vec v}){v^i}
	+\frac{4}{9}({\vec a}''\!\cdot\! {\vec v}){a^i}
	+\frac{2}{3}a^2{a^i}'
	+\frac{2}{3}({\vec a}' \!\cdot\! {\vec a}){a^i}
 \]  ,~~~
\label{F Galtsov 6d EM LO,NLO}
\eea
from which we also find an expression for the emitted power,
\bea
P_{Galt'sov} &=& -{\vec v} \cdot {\vec F}_{Galt'sov} = \nonumber\\
&=&	\frac{4}{45}({\vec v} \cdot {\vec a}''')
	+\frac{4}{15} {\vec v}^2 ({\vec v} \cdot {\vec a}''')
	+\frac{4}{3} ({\vec v}\!\cdot\! {\vec a}) ({\vec v}\!\cdot\! {\vec a}'')
	+\frac{8}{9}({\vec v} \!\cdot\! {\vec a}')^2
\nonumber\\ &&
	+\frac{2}{3} {\vec a}^2 ({\vec a}' \cdot {\vec v})
	+\frac{2}{3}({\vec a}' \!\cdot\! {\vec a}) ({\vec v} \cdot {\vec a}).
\label{P Galtsov 6d EM LO,NLO}
\eea
For our RR force calculation on a point charge $q$ along a path $\vec{x}(t)$, we rewrite the action
\bea
\hS_{EM}=\int\!dt\,\hat{L}_{EM}, \,\,\,\,\,\,\,\,\,\,\,\,\,\,\hat{L}_{EM}=\hat{L}_{EM}^S+\hat{L}_{EM}^V,
\eea
as a PN series expansion.
With (\ref{I EM scalar multipoles},\ref{J EM vector multipoles},\ref{S EM multipoles}, \ref{Bessel J series2}) we find for every (even) dimension
\bea
\hat{L}_{EM}^S=\,q^2\sum_{L}\frac{(-)^{\ell+\frac{\hd+1}{2}}(2\ell+\hd)!!}{\ell(\ell+\hd) \hd!! } 	\nonumber\\
	 &\!\!\!\!\!\!\!\!\!\!\!\!\!\!\!\!\!\!\!\!\!\!\!\!\!\!\!\!\!\!\!\!\!\!\!\!\!\!\!\!\!\!\!\!\!\!\!\!\!\!\!\!\!\!\!\!\!\!\!\!
	\cdot\,
	\sum_{\hat{p}=0}^\infty  \frac{\d_t^{2\hat{p}}}{(2\hat{p})!!(2\ell+2\hat{p}+\hd)!!}
		 \frac{\delta}{\delta x^{i}} \[ (2\hat{p}+\ell+\hd)r^{2\hat{p}}x_L
		-\d_t \( r^{2\hat{p}}x_L \vec{v} \cdot \vec{r}  \)   \] \hat{x}^{i}  \,\,\,\,\,\,\,\,\,\,\,\,\,\,\,\,\,\,		 \nonumber\\
	&\!\!\!\!\!\!\!\!\!\!\!\!\!\!\!\!\!\!\!\!\!\!\!\!\!\!\!\!\!\!\!\!\!\!\!\!\!\!\!\!\!\!\!\!\!\!\!\!\!\!
	\cdot\,
	\d_t^{2\ell+\hd}\!
	\sum_{p=0}^\infty \frac{ \d_t^{2p}}{(2p)!!(2\ell+2p+\hd)!!}
		 \[ (2p+\ell+\hd)r^{2p} x^L
		-\d_t \( r^{2p} x^L \vec{v} \cdot \vec{r}  \)   \] ^{STF} ,
		\,\,\,\,\,\,\,\,\,\,\,\,\,\,\,\,
	\nonumber\\
\eea
for the scalar part and
\bea
\hat{L}_{EM}^V=\,q^2\sum_{L}
\frac	{(-)^{\ell+\frac{\hd+1}{2}} \ell^2 (2\ell+\hd)!! }
		{(\hd-1)!! (\ell+1)(\ell+\hd-1)}
	\sum_{\hat{p}=0}^\infty &	 \frac{\d_t^{2\hat{p}}}{(2\hat{p})!!(2\ell+2\hat{p}+\hd)!!}
	\hat{x}^{i}\frac{\delta}{\delta x^{i}} \[ r^{2\hat{p}}(\vec{r} \wedge \vec{v})^{(k_\ell}x^{L-1)} \]
	\,\,\,\,\,\,\,\,\,\,\,\,\,\,	\nonumber\\
	\cdot\,
	\d_t^{2\ell+\hd}
	\sum_{p=0}^\infty &	 \frac{\d_t^{2p}}{(2p)!!(2\ell+2p+\hd)!!}
		 \[ r^{2p}(\vec{r} \wedge \vec{v})^{(k_\ell} x^{L-1)} \] ^{STF}\!\!,
		\,\,\,\,\,\,\,\,\,\,\,\,\,\,\,\,
\label{EM lagrangian}
\eea
for the vector part.\\
Similarly to the scalar RR computation, we integrate by parts to move the $2p$ (or $2p+1$) time derivatives from the $\hat{x}^L$ multipoles to the $x^L$ multipoles, and use the EOM (\ref{scalar EL})  found from variation with respect to $\hat{x}^j$.
We thus find the leading RR force, arising from the electric dipole term ($\ell=1,p=\hat{p}=0$, sources $\rho,\hat{\rho}$) at every (even) dimension $d$ to be
\bea
\vec{F}^{(d)}_{LO} = q^2\frac{(-)^{\frac{d}{2}}(d-2)}{(d-1)!! (d-3)!!} \d_t^{d-1}\vec{x}~.
\label{F EM LO any dimension}
\eea
The term for $d=4$ of course matches ALD; the term for $d=6$ is recorded in table \ref{table:EM leading scalar}, and matches that expected from Galt'sov (\ref{F Galtsov 6d EM LO,NLO}).
In fact, in every even dimension this expression matches exactly the leading order PN ($\frac{v}{c}<<1$) which can be derived from Kazinski, Lyakhovich \& Sharapov (\cite{Kazinski:2002mp}) eq. (26),(30), found in a very different method (\cite{Galtsov:2004qz,Kazinski:2004qq}).
We find the exact match to significantly support the validity of both our method and Kazinski et al's.\\
The next-to-leading-order includes 5 contributions to the scalar sector, summarized for $6d$ in table \ref{table:EM next to leading scalar}, as well as the leading vector contribution (table \ref{table:EM next to leading vector}).
Their sum is identical with Galt'sov's result (\ref{F Galtsov 6d EM LO,NLO})\footnote{Kosyakov \cite{Kosyakov:2008wa} gives a result for the power at $d=6$ that seems to agree at LO but disagree at NLO with our and with Galt'sov's result.}.
We have also tested these expressions using Mathematica Code\cite{Code} for dimensions (8,10,12,14,16), finding in every case non-trivial cancellations of all non-physical terms (the terms involving explicit position coordinates, breaking translation invariance).\\
We remark also that at $d=2$ the only STF tensor is $x$ ($\ell=1$), and we reproduce the expected null result for the action, radiation and self-force identically to any order.

\begin{table}[h!]
  \centering \caption{Leading order contribution to the $6d$ EM self-force (only electric $\eps=+$)}
\begin{center}
\begin{tabular}{cccccccc}
  \hline
  $\ell$ $p$ $\hat{p}$ & src & $\hat{L}/q^2$ & $F^j/q^2$ \\  \hline
  1 0 0 & $\rho$ $\hat\rho$
			& $-\frac{4}{45} \hat{x}^{i}\d_t^5x_{i}$
			& $-\frac{4}{45}\dddot{a}^j$\\  \hline
  \label{table:EM leading scalar}
\end{tabular}
\end{center}
\end{table}

\begin{table}[h!]
  \centering \caption{Next-to-Leading order contribution to the $6d$ EM self-force, scalar (electric $\eps=+$) sector}
\begin{center}
\begin{tabular}{cccccccc}
  \hline
  $\ell$ $p$ $\hat{p}$ & src & $\hat{L}/q^2$ & $F^j/q^2$ \\  \hline
  2 0 0 & $\rho$ $\hat\rho$
	& $\frac{1}{252}\hat{x}^{j} \frac{\delta}{\delta x^{j}}[x_i x_k]\d_t^7[x^i x^k-\frac{1}{5} x^2 \delta^{ik}]$
	& $\frac{1}{126}[x_i \d_t^7(x^i x^j)-\frac{1}{5}x^j \d_t^7 x^2]$	\vspace{1mm} \\  \hline
  1 1 0 & $\rho$ $\hat\rho$
	& $-\frac{1}{105}\hat{x}^{j} \frac{\delta}{\delta x^{j}}[x^2 x^i] \d_t^7 x_i$
	& $-\frac{1}{105}[x^2 \d_t^7 x^j+2x^j x_i \d_t^7 x^i]$	\vspace{1mm} \\  \hline
  1 0 1 & $\rho$ $\hat\rho$
	& $-\frac{1}{105}\hat{x}_i \d_t^7(x^i x^2)$
	& $-\frac{1}{105}\d_t^7(x^2 x^j)$	\vspace{1mm} \\  \hline
  1 0 0 & $j_r$ $\hat\rho$
	& $\frac{1}{45} \hat{x}^i \d_t^6[v_k x^k x_i]$
	& $\frac{1}{45}\d_t^6(x^j x_i v^i)$	\vspace{1mm} \\  \hline
  1 0 0 & $\rho$ $\hat{j}_r$
	& $-\frac{1}{45}\hat{x}^{j} \frac{\delta}{\delta x^{j}}[v_k x^k x^i]\d_t^6x_i$
	& $-\frac{1}{45}[x_i v^i \d_t^6x^j + v^j x_i \d_t^6  x^i
		-\frac{d}{dt}(x^j x^i \d_t^6x_i)]$	\vspace{1mm} \\  \hline
  \label{table:EM next to leading scalar}
\end{tabular}
\end{center}
\end{table}

\begin{table}[h!]
  \centering \caption{Next-to-Leading order contribution to the $6d$ EM self-force, from vector (magnetic $\eps=-$) sector}
\begin{center}
\begin{tabular}{cccccc}
  \hline
  $\ell$ $p$ $\hat{p}$ & $\hat{L}/q^2$ & $F^j/q^2$ \\  \hline
  1 0 0 & $-\frac{1}{90}\hat{x}^{i} \frac{\delta}{\delta x^{i}}[r_j v_k] \d_t^5(r^j v^k - r^k v^j)$
		& $-\frac{1}{90} \( 2v_i \d_t^5[x^j v^i - x^i v^j]+x_i \d_t^5[x^j a^i- x^i a^j] \)$
	\vspace{1mm} \\  \hline
  \label{table:EM next to leading vector}
\end{tabular}
\end{center}
\end{table}

{\bf Dissipated power}.
Similarly to (\ref{radiated energy scalar definition}), we compute the power of the RR force on the accelerating charge, now using (\ref{S EM multipoles}):
\bea
P_{RR} &=&-\vec{v}\cdot \vec{F}=-\frac{d x^{i}}{dt} \left.\frac{\delta \hat{L}}{\delta x^{i}}\right|_{\hat{\vec{x}} \to \vec{x}} \nonumber\\
&=&\!\!\sum_{L}\!\!\!\frac{(-)^{\ell+\frac{d}{2}}}{ (2\ell+\hd)!!\hd!!}\!\!
	\left. \[
		\frac{\ell+\hd}{\ell} \frac{d x^{i}}{dt}\frac{\delta Q_{(E)}^L}{d x^{i}} \!\cdot\! \d_t^{2\ell+\hd} Q^{(E)}_L
		+ \frac{\ell^2 \hd}{(\ell\!+\!1)(\ell\!+\!\hd\!-\!1)}\frac{d x^{i}}{dt}\frac{\delta Q_{(M)}^L}{d x^{i}}\!\cdot\! \d_t^{2\ell+\hd}Q^{(M)}_L
	 \]  \right|_{\hat{\vec{x}} \to \vec{x}}	\!\!\!\!\!\!\!\!\!\!\! .~~~~~~~\, \nonumber\\
\label{radiated energy EM definition}
\eea
The time-averaged power is found using
\bea
\int\!\!{dt} \frac{d x^{i}}{dt} \frac{\del Q_{(E)}^L}{\del x^{i}} = \int\!\!{dt} \frac{d Q_{(E)}^L}{dt}
	\,\,\,\,,\,\,\,\,
\int\!\!{dt} \frac{d x^{i}}{dt} \frac{\delta Q_{(M)}^L}{\delta x^{i}} = \int\!\!{dt} \frac{d Q_{(M)}^L}{dt} \, \, ,
\label{EM EL as full derivative}
\eea
followed by $\ell+\frac{\hd-1}{2}$ integrations by parts (and recalling $\hd$ is odd), to be
\bea
\label{radiated energy EM multipoles}
<\!\!P_{RR}\!\!>&=& \sum_{L}\frac{1}{ (2\ell+\hd)!!\hd!!}
	\left<
		\REone (\d_t^{\ell+\frac{\hd+1}{2}}Q_{(E)}^L)^2
		+ \RMone \frac{D_\ell (\hd+1,1)}{D_\ell (\hd+1,0)}  (\d_t^{\ell+\frac{\hd+1}{2}}Q_{(M)}^L)^2
	\right>	\\
&\!\!\!\!\!\!\!\!\!\!\!\!\!\!\!\!\!\!\!\!\!\!\!\!\!\!\!=&\!\!\!\!\!\!\!\!\!\!\!\!\!\!\!\!
	\!\sum_{L}\!\!\frac{(\ell\!+\!\hd)}{\ell (2\ell\!+\!\hd)!!\hd!!}
		\left<\!\! \(\d_t^{\ell+\frac{\hd+1}{2}}\!Q_{(E)}^L	 \) ^2 \right>
	\!\!+\!\! \sum_{L}\!\!\frac{\ell^2}{(\ell\!+\!1)(\ell\!+\!\hd\!-\!1)(2\ell\!+\!\hd)!!(\hd\!-\!1)!!}
		\left<\!\! \( \d_t^{\ell+\frac{\hd+1}{2}}\!Q_{(M)}^L	 \) ^2 \right>
	\nonumber \\
&\!\!\!\!\!\!\!\!\!\!\!\!\!\!\!\!\!\!\!\!\!\!\!\!\!\!\!=&\!\!\!\!\!\!\!\!\!\!\!\!\!\!\!\!
P_{rad} ~. \nonumber
\eea
In the 4$d$ case, we recognize this result as Ross' eq.(52) \cite{RossMultipoles} (with a $4\pi$ normalization factor, and re-introducing the $\frac{1}{\ell!}$ factor for comparison, see Appendix \ref{app:Multi-index summation convention}).

\section{Summary of results}
\label{section:Summary of results}

In this section we summarize the essential definitions and main results obtained in the paper.
We use bessel-like functions defined (\ref{Bessel J series2}) as
\be
\tilde{j}_\alpha := \Gm(\alpha+1) 2^{\alpha} \frac{J_\alpha(x)}{x^\alpha}\, = 1 \, + \, \ldots \, ,
\ee
where $J_\alpha(x)$ is the Bessel function of the first kind. $x^{TF}_L \, = \,\left[ x^{k_1} \, \ldots \, x^{k_{\ell}}\right]^{TF}$ are trace-free tensor products of spatial position vectors (\ref{nL STF}), using the summation conventions of (\ref{app:Multi-index summation convention}).
We write the radiation reaction (RR) effective action $\hat{S}$ in terms of the system's multipoles $Q^L$ and their doubled counterparts $\hat{Q} =  \frac{\delta Q}{\delta \rho} \hat{\rho}$ (\ref{sum: doubled multipoles}), and find the EOM by variation of $\hat{S}$ with respect to hatted fields (\ref{ctp intro}).\\

For a massless scalar field coupled to sources
\bea
S_\Phi= \frac{1}{2 \Omd\, G} \int\! (\d_\mu \Phi)^2 d^d x  - \int\!\! \rho \, \Phi d^d x\, , \,\,\,\,\,\,\,\,
\label{Scalar Action sum}
\eea
we find the radiative multipoles to be
\be
Q_L = \int d^D x\, \tilde{j}_{\ell+\frac{\hat{d}}{2}}(ir\d_t)\, x^{TF}_L\, \rho(\vec{r},t) ~ .
\label{I Phi scalar multipoles sum}
\ee
In even-dimensional spacetime, outgoing radiation is given by
\bea
\Phi(\vec{r},t) = \frac{G}{\hd!!} r^{-\frac{\hd+1}{2}}
\sum_L n^L \d_t^{\ell+\frac{\hd-1}{2}} Q_L(t-r) ~~  \, , ~
\label{radiation Phi sum}
\eea
and the power dissipated through it is
\bea
\dot{E}=\sum_{L} \frac{G}{\hd!!(2\ell+\hd)!!} \left<(\d_t^{\ell+\frac{\hd+1}{2}}Q^L )^2\right>
= \sum_{L} \frac{G}{\ell! \hd!! (2\ell+\hd)!!} \left<(\d_t^{\ell+\frac{\hd+1}{2}}Q_{k_1 k_2 \cdots k_\ell}^{STF})^2\right>\,.~~~~
\label{radiated energy scalar multipoles sum}
\eea
Elimination of system \& radiation zone fields gives the radiation reaction (RR) effective action
\bea
\hS_\Phi=
G\int\!\!{dt}\sum_{L} \frac{(-)^{\ell+\frac{\hd+1}{2}}}{\hd!!(2\ell+\hd)!!}\hat{Q}^L \d_t^{2\ell+\hd}Q_L~ .
\label{S Phi scalar multipoles}
\eea
In odd-dimensional spacetime, frequency domain results are similar to those of the even dimensional case. In the time domain, outgoing radiation is given by
\bea
\Phi(\vec{r},t) = \frac{- G}{ \sqrt{2 \pi} \hd!!} r^{-\frac{\hd+1}{2}}
\sum_L n^L \d_t^{\ell+\frac{\hd}{2}} \int^{t-r}_{-\infty} \frac{Q_L(t')}{\left| t-r-t' \right|^{1/2}} \, dt' ~~  \, , ~
\label{radiation Phi sum odd}
\eea
and the RR effective action by
\bea
\hS_\Phi=
G\!\!\int_{-\infty}^{\infty}\!\!\!\!\!\!\!\!{dt}\sum_{L}\! \frac{(-)^{\ell+\frac{\hd}{2}}}{\hd!!(2\ell+\hd)!!} \, \hat{Q}^L(t)
\[
	\(\frac{1}{2}H(2\ell+\hd)-H(\ell+\frac{\hd}{2}) \) \d_t^{2\ell+\hd}Q_L(t) \right. \nonumber\\
\left. \left.
	-\int_{-\infty}^{t}\!\!\!\! dt' \( \frac{1}{t-t'} \d_{t'}^{2\ell+\hd}Q_L(t') \) \right|_{regularized}
\]~ , ~~~~~~~~~~~~~~~~~~~~~~~~~~
\label{S Phi scalar multipoles odd dimension sum}
\eea
where the regularization is discussed in Sec. \ref{Odd spacetime dimensions}. The coefficient of the local term in (\ref{S Phi scalar multipoles odd dimension sum}) above is not universal but depends on the short-distance details of the system.

For an EM field coupled to sources
\bea
\label{MaxwellEMAction sum}
S= -\frac{1}{4 \Omd} \int\!\! F_{\mu \nu} F^{\mu \nu} r^{\hat{d}+1} d^d x  - \int\!\! A_{\mu} J^{\mu}  r^{\hat{d}+1} d^d x \, ,
\eea
we find the following electric and magnetic multipoles
\bea
Q^L_{(E)}&=&\frac{1}{\ell+\hd}\int\!\!{d^D x} x^L_{TF}
	 \[ \frac{1}{r^{\ell+\hd-1}} \( r^{\ell+\hd}\tilde{j}_{\alpha}(ir \d_t) \) ' \rho(\vec{x})
	-\tilde{j}_{\alpha}(ir \d_t)\, \d_t \vec{J}(\vec{x}) \cdot \vec{x} \,  \] , \nonumber \\
 Q^{L \, \aleph}_{(M)}&=& \int\!\!{d^D x} \, \, \tilde{j}_{\ell + \frac{\hat{d}-1}{2}}(ir\del_t)  \[ \epsilon^{k_\ell \, \aleph}_{\, a \, b} \, r^a \, J^b  x^{L-1} \] ^{STF} \, ,
\eea
where $\aleph$ is an antisymmetric multi-index (\ref{x L aleph Omega}). Radiation in even spacetime dimensions is given by
\bea
A_\eps(\vec{r},t) =\frac{1}{\hd!!} r^{-\frac{\hd+1}{2}}
\sum_L R^\eps_1 n^L \d_t^{\ell+\frac{\hd-1}{2}} Q^\eps_L(t-r) ~ \, , ~
\label{radiation A sum}
\eea
where $\eps$ is $(+)$ for the electric sector and $(- , \aleph)$ for the magnetic sectors, and
\bea
R^+_1 =\frac{\ell+\hat{d}}{\ell} ~~~;~~~R^-_1 =\frac{\ell}{\ell+\hat{d}}~~.
\eea
The RR effective action in even spacetime dimensions is
\bea
\hS_{EM}=\int\!\!{dt}\!\sum_{L}\frac{(-)^{\ell+\frac{\hd+1}{2}}}{\hd!! (2\ell+\hd)!!}
	 \[
		\frac{\ell+\hd}{\ell} \hat{Q}^{(E)}_L \cdot \d_t^{2\ell+\hd}Q_{(E)}^L
	+
		\frac{\ell^2 \hd}{(\ell\!+\!1)(\ell\!+\!\hd\!-\!1)} \hat{Q}^{(M)}_L \cdot \d_t^{2\ell+\hd}Q_{(M)}^L
	 \] \!\! \, \, \,  .~~~~ \nonumber \\
\label{S EM multipoles time sum}
\eea
The case of the EM field in non-even (and in particular odd) spacetime dimensions is treated in a similar manner to the scalar case, and similar non-local ``tail" expressions appear:
\bea
\hS_{EM}&=&\int\!\!{dt}\!\sum_{L}\frac{(-)^{\ell+\frac{\hd+1}{2}}}{\hd!! (2\ell+\hd)!!}
	 \[
		\frac{\ell+\hd}{\ell} S^{(E)}(t)
	+
		\frac{\ell^2 \hd}{(\ell\!+\!1)(\ell\!+\!\hd\!-\!1)} S^{(M)}(t)
	 \] \!\! ,~~~~
\label{S EM multipoles odd dimension 3}
\\  \nonumber\\ 
S^{(E/M)}(t) &=& \hat{Q}^{(E/M)}_L(t)
\[
	\(\frac{1}{2}H(2\ell+\hd)-H(\ell+\frac{\hd}{2}) \) \d_t^{2\ell+\hd}Q_{(E/M)}^L(t)  \right. \nonumber\\ 
&& \left. \left. ~~~~~~~~~~~~~~
	-\int_{-\infty}^{t}\!\!\!\! dt' \( \frac{1}{t-t'} \d_{t'}^{2\ell+\hd}Q_{(E/M)}^L(t') \) \right|_{regularized}
\]~ . ~
\label{S EM multipoles odd dimension 4}
\eea

\section{Discussion}
\label{section:Discussion}

In this paper we formulated an EFT describing radiative effects in scalar and EM theories in general spacetime dimensions and applied it to solve for the radiation and radiation reaction effective action in these cases, thereby generalizing the 4$d$ treatment of paper I.
We found that the method devised there \emph{naturally} generalizes to higher dimensions, providing new results even in these linear, well studied theories (see section \ref{section:Summary of results}) and laying a solid foundation for the study of such effects in higher dimensional GR.

Some dimension-dependent issues that need to be handled with care appeared.
One of them is the tail effect in odd spacetime dimensions (and, formally, in all non-even $d$), which is due to indirect propagation - propagation not restricted to the lightcone.
We found that while frequency domain analyses of any spacetime dimensions are similar, time domain results are substantially different.
From our analysis' point of view, the difference is all due to different analytic properties of the fields and effective actions in the complex frequency plane.
It appears only when transforming the results into the time domain.
In particular, we find that in odd $d$ the RR effective action is composed of a nonlocal part which contains all the dissipative effects, and a local conservative part; while in even $d$ the RR effective action contains only a local part which is purely dissipative.

We remark that there has been debate \cite{Kosyakov:1999np, Galtsov:2001iv, Kazinski:2002mp, Galtsov:2004qz, Kazinski:2004qq, Galakhov:2007my, Kosyakov:2008wa, Shuryak:2011tt} over the very possibility of defining and regularizing self-force and radiation-reaction even in general even dimensions.
We hope our independent method and results for scalar and EM fields in any dimension help shed new light on the matter.

Another important issue appeared when treating fields with nonzero spin, namely the issue of having multiple fields, and the associated question of gauges.
Here these arose in the case of the EM field of spin $1$.
One of the main ideas of our method is the use of gauge-invariant spherical fields and the reduction of the problem to $1d$.
This was done with a vector spherical harmonic decomposition.
In 4$d$, the electric field (which behaves like a scalar on the sphere) and magnetic field (which behaves like a vector on the sphere) are very similar - one is a vector field and the other an axial vector field.
However, the magnetic field is more generally a two-form field.
This is more visible when working in $d > 4$, where for example one obtains essentially different multipoles for these two sectors, which live in different representations of the rotation group.
Going to higher spins in higher dimensions, more sectors appear - a rank-2 tensor sector in GR, etc.

The \emph{main} complication that arises in the gravitational case is the theory's nonlinearity.
Although for the leading order the linearized theory suffices, just as in 4$d$, going to higher order one would need to include nonlinear interactions - a +1PN correction from first system zone nonlinearities, +1.5PN from first radiation zone nonlinearities, and so on.
Also, there is a PN order hierarchy between the different sectors: only the scalar sector is needed for leading order, for next-to-leading (+1PN) order the vector sector must also be accounted for, and at next-to-next-to-leading (+2PN) order the tensor sector also begins to contribute.
We treat the general $d$ gravitational 2-body problem and its non-linearities separately in \cite{BirnholtzHadar2015a}.

It would be interesting to use and enhance our method in the study of higher order radiation zone effects (first in $4d$).
For example, one can replace the Bessel functions in our propagators with Regge-Wheeler/Zerilli functions (solutions for gravitational perturbations around non-rotating black holes) and compare results for an observable - e.g., emitted radiation - to those obtained by the usual EFT sum of diagrams for scattering of waves off the total mass of the system.
This could both improve the EFT and give insight on the problem from a new angle.

\subsection*{Acknowledgments}
We thank the organizers of the 16th CAPRA meeting hosted in Dublin in July 2013 during which this work was initiated.
We thank Chad Galley, Tomer Shacham and Barry Wardell for helpful discussions.
We especially thank Barak Kol for many very useful discussions and encouragement, and for comments on the manuscript.
This research was supported by the Israel Science Foundation grant no. 812/11, and was part of the Einstein Research Project "Gravitation and High Energy Physics", which is funded by the Einstein Foundation Berlin.
O.~B. was partly supported by an ERC Advanced Grant to T. Piran.

\appendix

\section{Solution without zone separation}
\label{no zones}
\subsection{Scalar field}
\label{no zones scalar}
Looking back at the field equation for a scalar field $\Phi$ (\ref{Scalar Field Equation}) using the spherical decomposition (\ref{decomposition of scalar field},\ref{decomposition of scalar field and source}, \ref{Nld},\ref{scalar inverse source},\ref{scalar action source}), we see that it can be solved using the propagator (\ref{Phi propagator scalar}), yielding ($r\!>\!r'$)
\bea
\Phi_\w(\vec{r}) &=& \sum_L \! x^L \Phi_{Lw}(r)
	= \sum_L \! x^L \!\!\! \int \!\! dr' G^\Phi_{ret} (r,r') \rho^\Phi_{L\w} (r') \nonumber\\
&=&
-iG \sum_L \! \w^{2\ell+\hd}  \Mld \, x^L \tldhap(\w r)
	\! \int \!\! d^Dx'
	\tldja(\w r') x'_L \rho_\w(\vec{r}\,')
	 ~ .
\label{Phi scalar no zones}
\eea
Restricting to even integer spacetimes $d$ (odd $\hd$), the self-force arises only from the time-asymmetric part of the propagator.
Since $\tldya$ contains only odd powers of $\w$ and $\tldja$ only even powers, by expanding the spherical Bessel functions as series we find the time-asymmetric propagator is given by
\bea
G_\w^{odd}(\vec{r},\vec{r}\,') &=&
-iG \sum_L \! \w^{2\ell+\hd}  \Mld \, x^L \tldja(\w r) \tldja(\w r') x'_L 	\nonumber\\
&=&
G \sum_L \! \frac{(-)^{\ell+\frac{\hd-1}{2}} (2\ell+\hd)!!}{\hd!!}
\sum_{p=0}^{\infty} \sum_{\hat{p}=0}^{\infty}\!\!
\frac{ (-i\w)^{2\ell+2p+2\hat{p}+\hd} r^{2\hat{p}} x^L r'^{2p} x'_L }
{ (2\hat{p})!!(2\ell+2\hat{p}+\hd)!! (2p)!!(2\ell+2p+\hd)!!} ,~~~~~~
\label{explicit odd green's function in frequency domain no zones}
\eea
and
\bea
\Phi_\w(\vec{r}) =
\int \!\! d^Dx'  G_\w^{odd}(\vec{r},\vec{r}\,') \rho_\w(\vec{r}\,') ~.~
\label{solution r'<r no zones}
\eea
Here we find it more direct to use the $\{r,r'\}$ basis rather than the Keldysh basis. The lagrangian for the self-interactions of a scalar-charged point particle $q$ described by (\ref{scalar point source}) is given in the by
\bea
L_\w&=&\int \!\! d^Dx \hat\rho_w(\vec{r}) \Phi_w(\vec{r}) = \int \!\! d^Dx d^Dx' \hat\rho_w(\vec{r}) G^{odd}_w(\vec{r},\vec{r} ') \rho_w(\vec{r})=
\label{Lagrangian no zones}
\\
&=&
G q^2 \! \sum_L \! \frac{(-)^{\ell+\frac{\hd-1}{2}} (2\ell+\hd)!!}{\hd!!} \!
\sum_{p=0}^{\infty} \sum_{\hat{p}=0}^{\infty} \sum_{s=0}^{\infty} \sum_{\hat{s}=0}^{\infty} \!\! C_{\ell,\hd}^{p \hat{p} s \hat{s}}
(-i\w)^{2\ell+2p+2\hat{p}+\hd}
\( r^{2p} v^{2s} x^L \)
\!\!
\(  {r'}^{2\hat{p}} {v'}^{2\hat{s}} x'_L \)
\!\!,\!\!\!
\nonumber\\
&& C_{\ell,\hd}^{p \hat{p} s \hat{s}} =
\frac{ (2s-3)!! (2\hat{s}-3)!! }
{ (2\hat{p})!!(2\ell+2\hat{p}+\hd)!! (2p)!!(2\ell+2p+\hd)!! (2\hat{s})!! (2s)!!},
\eea
similarly to the one given by (\ref{scalar lagrangian}), and where an $\ell!^{-1}$ is implied by the summation convention.
The self-force on the particle is found using the Euler-Lagrange equation by first varying according to the difference in trajectories $r-r'$ (equivalent to the hatted trajectory of the Keldysh basis), then setting $r'\!\!\to\!\! r$; we see immediately that the force found in (\ref{scalar EL}) is recovered exactly.

\subsection{Electromagnetic self-force in the Lorentz Gauge}
\label{no zones EM lorentz}

We examine the EM action for the field $A^\mu$ in $d$ spacetime dimensions (\ref{MaxwellEMAction}), with a source $d$-current
\bea
j^{\mu}=(\rho,\vec{j}),
\label{EM source no zones}
\eea
\bea
\rho=q\delta^{(D)}(x-x_p(t)),
\label{EM source0 no zones}
\eea
\bea
\vec{j}=q\vec{v_p}\delta^{(D)}(x-x_p(t)) .
\label{EM source vector no zones}
\eea
Under the Lorenz gauge condition, the wave equation (\ref{Scalar Field Equation}) is replaced for the Electromagnetic field $A^{\mu}=(\phi,\vec{A})$ by
\bea
\Box A^{\mu}= \Omd j^\mu .
\label{EM EOM}
\eea
These $d$ independent equations are each of the form (\ref{Scalar Field Equation}), except for an opposite overall sign (and $G=1$).
We can thus use the same propagator (\ref{explicit odd green's function in frequency domain no zones}), and we solve the $d$ equations using the same method as in the scalar case: we fourier transform each equation along with its source, integrate using this propagator $G^{odd}$ in the frequency domain, and transform back to find the $d$ EM components, analogously to (\ref{solution r'<r no zones}):
\bea
A_\w^\mu(\vec{r}) = \int \!\! d^D x' G_\w^{odd}(\vec{r},\vec{r}') j_w^\mu(\vec{x}') .
\label{EM integral}
\eea
We note that for the scalar potential term $A^0=\phi$, this is just equation (\ref{solution r'<r no zones}) again, where the only difference from the scalar case is the simpler source term (\ref{EM source0 no zones}) instead of (\ref{scalar point source}), or in other words the absence of the $\gamma$ term.
This means we will find the same contributions to this potential term as we have found for the scalar potential, but without $s>0$ corrections.
For $A^k$ ($k=1..D$), solving (\ref{EM integral}) is similar to solving (\ref{solution r'<r no zones}), but with an additional $v'^k$ present in $j^k$:
\bea
\phi_\w(\vec{r}) &=&
q \sum_L \! \frac{(-)^{\ell+\frac{\hd+1}{2}} (2\ell+\hd)!!}{\hd!!}
\sum_{p=0}^{\infty} \sum_{\hat{p}=0}^{\infty}\!\!
\frac{ (-i\w)^{2\ell+2p+2\hat{p}+\hd} r^{2\hat{p}} x_L r_p^{2p} x^L_p}
{ (2\hat{p})!!(2\ell+2\hat{p}+\hd)!! (2p)!!(2\ell+2p+\hd)!!}~,~~~~\\
A^k_\w(\vec{r}) &=&
q \sum_L \! \frac{(-)^{\ell+\frac{\hd+1}{2}} (2\ell+\hd)!!}{\hd!!}
\sum_{p=0}^{\infty} \sum_{\hat{p}=0}^{\infty}\!\!
\frac{ (-i\w)^{2\ell+2p+2\hat{p}+\hd} r^{2\hat{p}} x_L r_p^{2p} x^L_p v_p^k}
{ (2\hat{p})!!(2\ell+2\hat{p}+\hd)!! (2p)!!(2\ell+2p+\hd)!!}~,~~~~
\eea
and an $\ell!^{-1}$ is implied by the summation convention.\\
We again solve order by order.
The leading and next-to-leading orders are recorded in tables \ref{table:vector field leading no zones},\ref{table:vector field next-to-leading no zones 0},\ref{table:vector field next-to-leading no zones vector}.\\

\begin{table}[h!]
  \centering \caption{Leading order contributions from $A^0$ and $A^k$}
\begin{center}
\begin{tabular}{cccccc}
  \hline
  $\ell$ & $p$ & $\hat{p}$ & & \\  \hline
  & & & $A^0/q=\phi/q$ & $F^j/q^2$	\\  \hline
  1 & 0 & 0
	& $ (-)^{\frac{d}{2}}\frac{1}{D!! \hd!!} x_{i}(x_{p}^{i})^{(D)}$
	& $ (-)^{\frac{d}{2}+1}\frac{1}{D!! \hd!!} (x_p^{j})^{(D)}$	\vspace{1mm} \\  \hline
  & & & $A^k/q$ & $F^j/q^2$ \\  \hline
  0 & 0 & 0
	& $ (-)^{\frac{d}{2}-1}\frac{1}{\hd!!^2} {v_p^k}^{(\hd)}$
	& $ (-)^{\frac{d}{2}}\frac{1}{\hd!!^2} (x_p^j)^{(D)}$	\vspace{1mm} \\  \hline
  \label{table:vector field leading no zones}
\end{tabular}
\end{center}
\end{table}

\begin{table}[h!]
  \centering \caption{Next-to-leading order contributions from $A^0$}
\begin{center}
\begin{tabular}{cccccc}
  \hline
  $\ell$ & $p$ & $\hat{p}$ & & \\  \hline
  & & & $A^0/q=\phi/q$  & $F^j/q^2$\\  \hline
  0 & 1 & 1
	& $ \frac{(-)^{\frac{d}{2}-1}}{ 4 D!!^2} x^2 (x_p^2)^{(d+1)}$
	& $ \frac{(-)^{\frac{d}{2}}}{2 D!!^2} x_p^j(x_p^2)^{(d+1)}$	\vspace{1mm} \\  \hline
  1 & 0 & 1
	& $ \frac{(-)^{\frac{d}{2}}} { 2\hd!!(d+1)!!}	x^2 x_i (x^i_p)^{(d+1)}$
	& $ \frac{(-)^{\frac{d}{2}-1}} { 2\hd!!(d+1)!!} \[ 2 x_p^j {x_p}_i(x_p^i)^{(d+1)} + x_p^2(x_p^j)^{(d+1)} \]$ 	 \vspace{1mm} \\  \hline
  1 & 1 & 0
	& $ \frac{(-)^{\frac{d}{2}}} { 2\hd!!(d+1)!!}	x_i (x_p^2 x^i_p)^{(d+1)}$
	& $ \frac{(-)^{\frac{d}{2}-1}} { 2\hd!!(d+1)!!} (x_p^j x_p^2)^{(d+1)}$	\vspace{1mm} \\  \hline
  2 & 0 & 0
	& $ \frac{(-)^{\frac{d}{2}-1}} { 2 \hd!! (d+1)!!}
		\[ x_i x_k (x^i_p x^k_p)^{(d+1)} - \frac{1}{D} x^2 (x^2_p)^{(d+1)} \]$
	& $ \frac{(-)^{\frac{d}{2}}} { \hd!! (d+1)!!}
		\[ {x_p}_i (x_{p}^{i} x_{p}^{j})^{(d+1)}-\frac{1}{D} x_p^j (x_{p}^2)^{(d+1)} \]$	\vspace{1mm} \\  \hline
  \label{table:vector field next-to-leading no zones 0}
\end{tabular}
\end{center}
\end{table}

\begin{table}[h!]
  \centering \caption{Next-to-leading order contributions from $A^k$}
\begin{center}
\begin{tabular}{cccccc}
  \hline
  $\ell$ & $p$ & $\hat{p}$ & $A^k/q$  & $F^j/q^2$\\  \hline
  0 & 0 & 1
	& $ \frac{(-)^{\frac{d}{2}-1}}{2 D!! \hd!!} x^2 (v_p^k)^{(D)} $
	& $ \frac{(-)^{\frac{d}{2}-1}}{2 D!! \hd!!}
		\[2x_p^j {v_p}_k {v_p^k}^{(D)}-x_p^2{v_p^j}^{(d)}
		-2(\vec{r}_p\cdot\vec{v}_p) {v_p^j}^{(D)} \]$ 	\vspace{1mm} \\  \hline
  0 & 1 & 0
	& $ \frac{(-)^{\frac{d}{2}-1}} { 2 D!! \hd!!} (x_p^{2} v_p^k)^{(D)} $
	& $ \frac{(-)^{\frac{d}{2}}} { 2 D!! \hd!!} (x_p^{2} v_p^j)^{(d)}$	\vspace{1mm} \\  \hline
  1 & 0 & 0
	& $ \frac{(-)^{\frac{d}{2}}}{ D!! \hd!!} x_i (x^i_p v_p^k)^{(D)} $
	& $ \frac{(-)^{\frac{d}{2}}}{ D!! \hd!!}
		\[{v_p}_k (v_p^k x_p^j)^{(D)} - {x_p}_i (v_p^j x_p^i)^{(d)}-{v_p}_i (v_p^j x_p^i)^{(D)}\]$	\vspace{1mm} \\  \hline
  \label{table:vector field next-to-leading no zones vector}
\end{tabular}
\end{center}
\end{table}

We derive the EM force from the Action for the particle trajectory in the EM field:
\bea
S=S_{worldline}+S_{EM}=-m\int\!\!d\tau-\int\!\!d\tau j_{\mu}A^{\mu}.
\label{Total Action EM}
\eea
When deriving the E-L equation, the first part introduces the $m\ddot{\vec{x}}$ term, while
\bea
S_{EM}=\int\!\!dt q(\vec{A}\cdot\vec{v}-\phi)=\int\!\!dtL_{EM}
\label{EM Action}
\eea
gives the EM force.
We thus find
\bea
F_{EM}^j=q(v_k\frac{\d A^k}{\d x_j}-\dot{A}^j-\frac{\d\phi}{\d x_j})=F_A^j+F_\phi^j
\label{EM Force from Action}
\eea
The leading and next-to-leading contributions to the self force are recorded in tables \ref{table:vector field leading no zones},\ref{table:vector field next-to-leading no zones 0},\ref{table:vector field next-to-leading no zones vector}.
Summed together, we find the two leading orders of the total self force on an EM charge $q$ in $d$ spacetime dimensions to be
\bea
\vec{F}_{LO}&=&
(-)^{\frac{d}{2}} \frac{D - 1}{(D-2)!! D!!} q^2 \vec{x}_p^{(D)}	~,
\\
\vec{F}^{d=4}_{NLO}&=&
	+q^2 \[
		\frac{2}{3}v^{2}_{p}\dot{\vec{a}}_{p}
		+ 2(\vec{v}_{p} \cdot \vec{a}_{p})\vec{a}_{p}
		+ \frac{2}{3}(\vec{v}_{p} \cdot \dot{\vec{a}}_{p})\vec{v}_{p}
		~\] .
\\
\vec{F}^{d=6}_{NLO}&=&
	-q^2\[
		\frac{2}{3}({\vec a}^2) \dot{\vec{a}}
		+\frac{2}{3}({\vec a}\cdot\dot{\vec a}) \vec{a}
		+\frac{8}{9}({\vec v}\cdot{\vec a}) \ddot{\vec{a}}
		\right. \nonumber\\ && ~~~~~~\left.
		+\frac{8}{9}({\vec v}\cdot\dot{\vec a}) \dot{\vec{a}}
		+\frac{4}{9}({\vec v}\cdot\ddot{\vec a}) {\vec{a}}
		+\frac{8}{45}({\vec v}^2) \dddot{\vec{a}}
		+\frac{4}{45}({\vec v}\cdot\dddot{\vec a}) {\vec{v}}
		~\] .
\eea
We see that the leading order matches ALD for $d=4$ and Galt'sov for $d=6$ (the numerical factors are $+\frac{2}{3}$ and $-\frac{4}{45}$, correspondingly), and that the next-to-leading order also exactly matches the expected ALD result for $d=4$, and Galt'sov's result for $d=6$ (\ref{F Galtsov 6d EM LO,NLO}) \cite{Galtsov:2007zz}.
We also record the emitted power for $d=6$ (LO \& NLO),
\bea
P^{d=6}&=&-\vec{v} \cdot \( \vec{F}^{d=6}_{LO} + \vec{F}^{d=6}_{NLO} \) \\
	&=& \!\! q^2 \!\! \[
		\frac{4}{45} ({\vec v}\!\cdot\!\dddot{\vec a})
		\!+\!\frac{2}{3} {\vec a}^2 ({\vec v}\!\cdot\!\dot{\vec a})
		\!+\!\frac{2}{3} ({\vec v}\!\cdot\!{\vec a}) ({\vec a}\!\cdot\!\dot{\vec a})
		\!+\!\frac{4}{3} ({\vec v}\!\cdot\!{\vec a})({\vec v}\!\cdot\!\ddot{\vec a})
		\!+\!\frac{8}{9} ({\vec v}\!\cdot\!\dot{\vec a})^2
		\!+\!\frac{4}{15} {\vec v}^2 ({\vec v}\!\cdot\!\dddot{\vec a})
		\]\!\!\!.~~~~~~~~\, \nonumber
\label{power 6d EM}
\eea
Code for computing the self-force for any even dimension is available on our webserver\cite{Code}.

\section{Useful definitions and conventions}
\label{app:defs}

In this appendix we collect several definitions and conventions used in this paper.

\subsection{Multi-index summation convention}
\label{app:Multi-index summation convention}

Multi-indices are denoted by upper case Latin letters
\be
 I \equiv I_\ell :=(i_1 \dots i_\ell) \, \, .
\ee
Here each $i_k=1,\ldots,D$ is an ordinary spatial index, and $\ell$ is the total number of indices.
We define a slightly modified summation convention for multi-indices by
\be
P_I\, Q_I := \sum_l P_{I_\ell}\, Q_{I_\ell} := \sum_\ell \frac{1}{\ell!} P_{i_1 \dots i_\ell}\, Q_{i_1 \dots i_\ell} \, \, ,
\ee
so repeated multi-indices are summed over as in the standard summation convention, but an additional division by $\ell!$ is implied.
When $\ell$ is unspecified the summation is over all $\ell$.\\
In addition a multi-index delta-like function is defined through
\be
\delta_{I_\ell J_\ell} := \ell!\, \delta_{i_1 j_1} \dots \delta_{i_\ell j_\ell} \, \, .
\ee
These definitions are such that factors of $\ell!$ are accounted for automatically.

\subsection{Normalizations of Bessel functions}
\label{app:Normalizations of Bessel functions}
We find it convenient to define an origin-biased normalization of Bessel functions.
We start with conventionally normalized solutions of the Bessel equation
\be
\[ \del_x^2 + \frac{1}{x} \del_x + 1 - \frac{\nu^2}{x^2} \] B_\nu(x)=0 \, \, ,
\ee
where $B \equiv \{J,Y,H^\pm\}$, namely $B$ represents Bessel (of the first or second kind) and Hankel functions, and $\nu$ denotes their order.
Given $\alpha$, we define
\be
\tilde{b}_\alpha := \Gm(\alpha+1) 2^{\alpha} \frac{B_\alpha(x)}{x^\alpha}\, \, ,
\ee
the ``origin normalized'' Bessel functions $\tilde{b}_\alpha$, which satisfy the equation (compare \ref{Modified Bessel equation})
\be
\[ \del_x^2 + \frac{2\alpha+1}{x}\del_x + 1\] \tilde{b}_\alpha(x) =0 ~.
\label{Modified Bessel equation2}
\ee
The purpose of this definition is to have a simple behavior of $\tilde{j}_\alpha$ in the vicinity of the origin $x=0$,
\be
\tilde{j}_\alpha (x) = 1 + \co\(x^2\) ~.
\ee
More precisely, the series expansion for $\tilde{j}_\alpha (x)$ around $x=0$ is given by the \emph{even} series
\be
\tilde{j}_\alpha (x)  = \sum_{p=0}^\infty \frac{(-)^p \, (2\alpha)!!}{(2p)!! (2p+2\alpha)!!} x^{2p} ~.
\label{Bessel J series2}
\ee
The asymptotic form of our solutions for $x\to\infty$ is best stated in terms of Hankel functions $\tilde{h}^\pm:= \tilde{j} \pm i \tilde{y}$
\be
\tilde{h}^\pm_\alpha(x) \sim (\mp i)^{\alpha+1/2} \frac{2^{\alpha+1/2} \Gm(\alpha+1)}{\sqrt{\pi}} \frac{e^{\pm i x}}{x^{\alpha+1/2}} \, \, .
\label{Bessel H asymptotic2}
\ee
Around $x=0$ the Bessel function of the second kind for $\al$ non-integer is
\bea
\tilde{y}_\alpha (x) &=&\frac{\Gm(\alpha+1) 2^{\alpha}}{x^\alpha}  \frac{\cos(\alpha\pi) J_\alpha (x) - J_{-\alpha}(x)}{sin(\alpha\pi)} \nonumber \\
&=&
\frac{\Gm(\alpha+1)2^\alpha}{sin(\alpha\pi)} \sum_{p=0}^\infty \frac{(-)^p}{(2p)!!} \frac{x^{2p}}{2^p}
	 \[  \frac{cos(\alpha\pi)}{2^\alpha \Gm(p+\alpha+1)} -\frac{2^\alpha x^{-2\alpha}}{\Gm(p-\alpha+1)}  \] ,
\label{Bessel Y series2}
\eea
while for integer $\al=n$, using (c.f. equation (9.1.11) of \cite{AS}) it is given by
\bea
\tilde{y}_n (x) &=&
	- \frac{2 (2n)!!}{\pi} \sum_{m=1}^{n} \frac{(2m-2)!!}{(2n-2m)!!} x^{-2m}
	+ \frac{2}{\pi} \ln \( \frac{x}{2} \) \tilde{j}_n(x)
\nonumber \\
&&
	-\frac{(2n)!!}{\pi} \sum_{k=0}^{\infty}
		\left[ \psi(k+1) + \psi(n+k+1) \right]
		\frac{(-)^{k}}{(2k)!!(2n+2k)!!} x^{2k}
\, \, ,
\label{Bessel Y series integer}
\eea
where $\psi$ is the digamma function, defined for integers using the Harmonic numbers $H(N)$ and the Euler-Mascheroni constant $\gm$:
\bea
\psi(N+1)=H(N)-\gm.
\eea



\begin{thebibliography}{99}


\bibitem{BirnholtzHadarKol2013a}
  O.~Birnholtz, S.~Hadar and B.~Kol,
  ``Theory of post-Newtonian radiation and reaction,''
  Phys.\ Rev.\ D {\bf 88}, 104037 (2013)
  [arXiv:1305.6930 [hep-th]].

\bibitem{GoldbergerRothstein1}
  W.~D.~Goldberger and I.~Z.~Rothstein,
 ``An effective field theory of gravity for extended objects,''
  Phys.\ Rev.\  D {\bf 73}, 104029 (2006) \\
  W.~D.~Goldberger,
  ``Les Houches lectures on effective field theories and gravitational radiation,''
  arXiv:hep-ph/0701129

\bibitem{GoldbergerRothstein2}
  W.~D.~Goldberger and I.~Z.~Rothstein,
  ``Dissipative effects in the worldline approach to black hole dynamics,''
  Phys.\ Rev.\ D {\bf 73}, 104030 (2006)
  [hep-th/0511133].

\bibitem{GoldbergerRoss}
  W.~D.~Goldberger, A.~Ross,
  ``Gravitational radiative corrections from effective field theory,''
  Phys.\ Rev.\  {\bf D81}, 124015 (2010).
  [arXiv:0912.4254 [gr-qc]].

\bibitem{GalleyTiglio}
  C.~R.~Galley and M.~Tiglio,
  ``Radiation reaction and gravitational waves in the effective field theory approach,''
  Phys.\ Rev.\ D {\bf 79}, 124027 (2009)
  [arXiv:0903.1122 [gr-qc]].

\bibitem{FoffaSturani4PNa}
  S.~Foffa and R.~Sturani,
  ``Tail terms in gravitational radiation reaction via effective field theory,''
  Phys.\ Rev.\ D {\bf 87}, 044056 (2013)
  [arXiv:1111.5488 [gr-qc]].

\bibitem{GalleyLeibovich}
  C.~R.~Galley and A.~K.~Leibovich,
  ``Radiation reaction at 3.5 post-Newtonian order in effective field theory,''
  Phys.\ Rev.\ D {\bf 86}, 044029 (2012)
  [arXiv:1205.3842 [gr-qc]].

\bibitem{GalleyNonConservative}
  C.~R.~Galley,
  ``The classical mechanics of non-conservative systems,''
  Phys.\ Rev.\ Lett.\  {\bf 110}, 174301 (2013)
  [arXiv:1210.2745 [gr-qc]].

\bibitem{Cardoso:2008gn}
  V.~Cardoso, O.~J.~C.~Dias and P.~Figueras,
  ``Gravitational radiation in $d>4$ from effective field theory,''
  Phys.\ Rev.\ D {\bf 78}, 105010 (2008)
  [arXiv:0807.2261 [hep-th]].


 \bibitem{Abraham}
  M.~Abraham,
  ``Theorie der Elektrizit\"{a}t", vol. II, Springer, Leipzig (1905).

 \bibitem{Dirac}
  P.~A.~M.~Dirac,
  ``Classical theory of radiating electrons,''
  Proc.\ Roy.\ Soc.\ Lond.\ A {\bf 167}, 148 (1938).

 \bibitem{JacksonALD}
 J. D. Jackson, Classical Electrodynamics, Third Edition (Wiley, New York, 1998).

\bibitem{Poisson:1999tv}
  E.~Poisson,
  ``An Introduction to the Lorentz-Dirac equation,''
  gr-qc/9912045.






\bibitem{Kosyakov:1999np}
  B.~P.~Kosyakov,
  ``Exact solutions of classical electrodynamics and the Yang-Mills-Wong theory in even-dimensional space-time,''
  Theor.\ Math.\ Phys.\  {\bf 119}, 493 (1999)
  [Teor.\ Mat.\ Fiz.\  {\bf 119}, 119 (1999)]
  [hep-th/0207217].

\bibitem{Galtsov:2001iv}
  D.~V.~Gal'tsov,
  ``Radiation reaction in various dimensions,''
  Phys.\ Rev.\ D {\bf 66}, 025016 (2002)
  [hep-th/0112110].

\bibitem{Kazinski:2002mp}
  P.~O.~Kazinski, S.~L.~Lyakhovich and A.~A.~Sharapov,
  ``Radiation reaction and renormalization in classical electrodynamics of point particle in any dimension,''
  Phys.\ Rev.\ D {\bf 66}, 025017 (2002)
  [hep-th/0201046].

\bibitem{Galtsov:2004qz}
  D.~V.~Gal'tsov and P.~Spirin,
  ``Radiation reaction reexamined: Bound momentum and Schott term,''
  Grav.\ Cosmol.\  {\bf 12}, 1 (2006)
  [hep-th/0405121].

\bibitem{Kazinski:2004qq}
  P.~O.~Kazinski, S.~L.~Lyakhovich and A.~A.~Sharapov,
  ``A Comment on `Radiation reaction reexamined: Bound momentum and Schott term by D.V. Gal'tsov and P. Spirin',''
  hep-th/0405287.

\bibitem{Galakhov:2007my}
  D.~Galakhov,
  ``Self-interaction and regularization of classical electrodynamics in higher dimensions,''
  JETP Lett.\  {\bf 87}, 452 (2008)
  [arXiv:0710.5688 [hep-th]].

\bibitem{Kosyakov:2007qc}
  B.~P.~Kosyakov,
  ``Introduction to the classical theory of particles and fields,''
  Berlin, Germany: Springer (2007) 479 p

\bibitem{Kosyakov:2008wa}
  B.~P.~Kosyakov,
  ``Electromagnetic radiation in even-dimensional spacetimes,''
  Int.\ J.\ Mod.\ Phys.\ A {\bf 23}, 4695 (2008)
  [arXiv:0803.3304 [hep-th]].

\bibitem{Shuryak:2011tt}
  E.~Shuryak, H.~-U.~Yee and I.~Zahed,
  ``Self-force and synchrotron radiation in odd space-time dimensions,''
  Phys.\ Rev.\ D {\bf 85}, 104007 (2012)
  [arXiv:1111.3894 [hep-th]].


\bibitem{Mironov:2007nk}
  A.~Mironov and A.~Morozov,
  ``Radiation beyond four space-time dimensions,''
  Theor.\ Math.\ Phys.\  {\bf 156}, 1209 (2008)
  [hep-th/0703097 [HEP-TH]].

\bibitem{Mironov:2007mv}
  A.~Mironov and A.~Morozov,
  ``On the problem of radiation friction beyond 4 and 6 dimensions,''
  Int.\ J.\ Mod.\ Phys.\ A {\bf 23}, 4677 (2008)
  [arXiv:0710.5676 [hep-th]].

\bibitem{Galtsov:2007zz}
  D.~V.~Gal'tsov and P.~A.~Spirin,
  ``Radiation reaction in curved even-dimensional spacetime,''
  Grav.\ Cosmol.\  {\bf 13}, 241 (2007)
  [arXiv:1012.3085 [gr-qc]].

\bibitem{Galtsov:2010cz}
  D.~Gal'tsov,
  ``Radiation Reaction and Energy-Momentum Conservation,''
  Fundam.\ Theor.\ Phys.\  {\bf 162}, 367 (2011)
  [arXiv:1012.2846 [gr-qc]].

\bibitem{AsninKol}
  V.~Asnin and B.~Kol,
  ``Dynamical versus auxiliary fields in gravitational waves around a black hole,''
  Class.\ Quant.\ Grav.\  {\bf 24}, 4915 (2007)
  [hep-th/0703283].

\bibitem{CTP}
  J.~S.~Schwinger,
  ``Brownian motion of a quantum oscillator,''
  J.\ Math.\ Phys.\  {\bf 2}, 407 (1961).\\
  K.~T.~Mahanthappa,
  ``Multiple production of photons in quantum electrodynamics,''
  Phys.\ Rev.\  {\bf 126}, 329 (1962).\\
  L.~V.~Keldysh,
  ``Diagram technique for nonequilibrium processes,''
  Zh.\ Eksp.\ Teor.\ Fiz.\  {\bf 47}, 1515 (1964)
  [Sov.\ Phys.\ JETP {\bf 20}, 1018 (1965)].


\bibitem{RubinOrdonez}
  M.~A.~Rubin and C.~R.~Ordonez,
  ``Eigenvalues and degeneracies for n-dimensional tensor spherical harmonics,''
  J.\ Math.\ Phys.\  {\bf 25}, 2888 (1984).
  M.~A.~Rubin and C.~R.~Ordonez,
  ``Symmetric Tensor Eigen Spectrum of the Laplacian on $n$ Spheres,''
  J.\ Math.\ Phys.\  {\bf 26}, 65 (1985).

\bibitem{Higuchi:1986wu}
  A.~Higuchi,
  ``Symmetric Tensor Spherical Harmonics on the $N$ Sphere and Their Application to the De Sitter Group SO($N$,1),''
  J.\ Math.\ Phys.\  {\bf 28}, 1553 (1987)
  [Erratum-ibid.\  {\bf 43}, 6385 (2002)].


\bibitem{Maxwell}
J.~C.~Maxwell,
  A treatise on electricity and magnetism, vol 1, Clarendon, Oxford, UK, p155ff (1873).\\
J.~C.~Maxwell,
  The scientific papers of James Clerk Maxwell.
  University Press, Cambridge (1890).

\bibitem{FryeEfthimiou}
  C.~Frye and C.~J.~Efthimiou,
  ``Spherical Harmonics in p Dimensions,"
  (2012)
  [arXiv:1205.3548 [math.CA]].

\bibitem{Kalf}
  H.~Kalf,
  ``On the expansion of a function in terms of spherical harmonics in arbitrary dimensions,"
  Bulletin of the Belgian Mathematical Society - Simon Stevin {\bf 2.4}, 361-380 (1995).

\bibitem{Applequist}
  J.~Applequist,
  ``Traceless cartesian tensor forms for spherical harmonic functions: new theorems and applications to electrostatics of dielectric media,"
  J. Phys. A: Math. Gen. {\bf 22}, 4303 (1989).

\bibitem{RossMultipoles}
  A.~Ross,
  ``Multipole expansion at the level of the action,''
  Phys.\ Rev.\ D {\bf 85}, 125033 (2012)
  [arXiv:1202.4750 [gr-qc]].

\bibitem{Blanchet:1989ki}
  L.~Blanchet and T.~Damour,
  ``Postnewtonian Generation Of Gravitational Waves,''
  Annales Poincare Phys.\ Theor.\  {\bf 50}, 377 (1989).

\bibitem{Damour:1990gj}
  T.~Damour and B.~R.~Iyer,
  ``Multipole analysis for electromagnetism and linearized gravity with irreducible cartesian tensors,''
  Phys.\ Rev.\ D {\bf 43}, 3259 (1991).

\bibitem{Detweiler:2002mi}
  S.~L.~Detweiler and B.~F.~Whiting,
  ``Selfforce via a Green's function decomposition,''
  Phys.\ Rev.\ D {\bf 67}, 024025 (2003)
  [gr-qc/0202086].


\bibitem{Balazs}
  N.~L.~Balazs,
  ``Wave Propagation in Even and Odd Dimensional Spaces,''
  Proc.\ Phys.\ Soc. A {\bf 68}, 521 (1955).

\bibitem{Ching:1995tj}
  E.~S.~C.~Ching, P.~T.~Leung, W.~M.~Suen and K.~Young,
  ``Wave propagation in gravitational systems: Late time behavior,''
  Phys.\ Rev.\ D {\bf 52}, 2118 (1995)
  [gr-qc/9507035].


\bibitem{Hestenes1}
  D.~Hestenes and G.~Sobczyk,
  Clifford Algebra to Geometric Calculus: A Unified Language for Mathematics and Physics
  (D.~Reidel Publishing Company, Dordrecht, 1984).

\bibitem{Hestenes2}
  D.~Hestenes,
  New Foundations for Classical Mechanics
  (Kluwer Academic Publishers, Dordrecht, 1990).

\bibitem{Baylis}
  W.~E.~Baylis, Electrodynamics, A Modern Geometric Approach
  (Birkh\"{a}user, Boston, 1999).

\bibitem{Doran}
  C.~Doran and A.~Lasenby,
  Geometric Algebra for Physicists
  (Cambridge University Press, 2003)

\bibitem{Hestenes3}
  D.~Hestenes,
  ``Oersted Medal Lecture 2002: Reforming the mathematical language of physics,"
  Am. \ J.\ Phys. 71, 104–121 (2003).

\bibitem{Rowland}
  D.~R.~Rowland,
  ``On the value of geometric algebra for spacetime analyses using an investigation of the form of the self-force on an accelerating charged particle as a case study,''
  Am.\ J.\ Phys 78, 187-194 (2010).

\bibitem{Chappell}
  J.~M.~Chappell, A.~Iqbal, D.~Abbott,
  ``A simplified approach to electromagnetism using geometric algebra," (2010)
  arXiv:1010.4947 [physics.ed-ph]


\bibitem{Code}
http://phys.huji.ac.il/\textasciitilde ofek/PNRR/

\bibitem{BirnholtzHadar2015a}
  O.~Birnholtz and S.~Hadar,
  ``Gravitational radiation-reaction in arbitrary dimension,''
  Phys.\ Rev.\ D {\bf 91}, no. 12, 124065 (2015)
  [arXiv:1501.06524 [gr-qc]].


\bibitem{AS}
  Abramowitz, M., and Stegun, I. (1972).
  \emph{Handbook of mathematical functions with formulas, graphs, and mathematical tables}.
  Dover Publications, Mineola, New York.

\end{thebibliography}
\end{document}